\documentclass[preprint,a4paper,nofootinbib]{revtex4-1}
\usepackage{textcomp}
\usepackage{amssymb,amsmath,MnSymbol,ifsym,latexsym,bm}
\usepackage{wrapfig,subfigure}
\usepackage[Euler]{upgreek}
\usepackage{longtable, supertabular}

\bibpunct[, ]{(}{)}{;}{a}{,}{,}

\usepackage[pdftex]{graphics}
\usepackage[pdftex,final]{graphicx}

\newcommand{\etas}{\upeta_\mathrm{S}}
\newcommand{\eetas}{\overline{\upeta}_\mathrm{S}}
\newcommand{\taupt}{\bm{\uptau}_\mathrm{p}}
\newcommand{\taup}[1]{\uptau_\mathrm{p,\,#1}}
\newcommand{\tauc}{\uptau_\mathrm{c}}
\newcommand{\etap}{\upeta_\mathrm{p}}
\newcommand{\etapo}{\upeta_{\mathrm{p},\,\mathrm{0}}}
\newcommand{\eetap}{\overline{\upeta}_\mathrm{p}}
\newcommand{\eetapo}{\overline{\upeta}_\mathrm{p,\,0}}
\newcommand{\eetapp}{\overline{\upeta}_\mathrm{p}^+}
\newcommand{\kB}{k_{\textsc{b}}}
\newcommand{\kBT}{k_{\textsc{b}} \,T}
\newcommand{\nkBT}{n\,k_{\textsc{b}} \,T}

\newcommand{\edot}{\dot{\varepsilon}}
\newcommand{\edote}{\dot{\varepsilon}_\mathrm{e}}

\newcommand{\Wi}{\mathrm{W\!i}}
\newcommand{\Wics}{\Wi_\mathrm{C}}
\newcommand{\Wisc}{\Wi_\Upsigma}
\newcommand{\Wie}{\mathrm{W\!i}_{\,\mathrm{e}}}
\newcommand{\Wip}{\mathrm{W\!i}^+}
\newcommand{\De}{\mathrm{De}}
\newcommand{\Dez}{\mathrm{De}_\mathrm{Z}}
\newcommand{\Oh}{\mathrm{Oh}}

\newcommand{\NK}{N_\textsc{k}}
\newcommand{\bK}{b_\textsc{k}}
\newcommand{\NIp}{N_{1,\,\mathrm{p}}}
\newcommand{\lamv}{\lambda_\mathrm{c}}
\newcommand{\lame}{\lambda_\mathrm{e}}
\newcommand{\lamo}{\lambda_\mathrm{0}}
\newcommand{\lamZ}{\lambda_\mathrm{Z}}
\newcommand{\phio}{\phi_\mathrm{0}}

\newcommand{\Mtens}{\bm{\mathrm{M}}}

\newcommand{\Mzz}{M_{zz}}
\newcommand{\Mrr}{M_{rr}}

\newcommand{\tr}{\mathrm{tr}}

\newcommand{\bQ}{\widetilde{\bm{Q}}}

\newcommand{\zetac}{\zeta_0}
\newcommand{\zetaZ}{\zeta_\mathrm{Z}}
\newcommand{\zetaR}{\zeta_\mathrm{R}}
\newcommand{\zetar}{\zeta_\mathrm{r}}
\newcommand{\zetas}{\zeta_\mathrm{s}}
\newcommand{\zetad}{\zeta_d}

\newcommand{\fs}{f_\mathrm{s}}
\newcommand{\ells}{\ell_\mathrm{s}}
\newcommand{\ds}{d_\mathrm{s}}

\newcommand{\cstar}{c^\ast}

\newcommand{\phistarc}{\phi_0^{\ast,\,c}}
\newcommand{\phistars}{\phi_0^{\ast,\,\Upsigma}}
\newcommand{\phidagrc}{\phi_0^{\dagger,\,c}}
\newcommand{\phidagrs}{\phi_0^{\dagger,\,\Upsigma}}

\newcommand{\Nd}{N_d}
\newcommand{\xih}{\xi_H}
\newcommand{\Nxi}{N_H}
\newcommand{\zetah}{\zeta_H}

\begin{document}

\title{Enhancement of coil--stretch hysteresis by self-concentration in extensional flows, and its implications for capillary thinning of liquid bridges of dilute polymer solutions}

\author{R. ~Prabhakar}
\email{prabhakar.ranganathan@monash.edu}
\affiliation{Department of Mechanical \& Aerospace Engineering, Monash University, Clayton,  AUSTRALIA}

\date{\today}

\begin{abstract}
The coil-stretch transition in extensional flows of viscoelastic dilute polymer solutions is known to be associated with a strong hysteresis in molecular conformations and rheo-optical properties. At infinite dilution, hysteresis is caused by the large difference in frictional drag coefficient between undeformed isotropic polymer coils and highly stretched conformations. At the low extension rates in the hysteresis regime, stretched molecules pervade larger volumes than equilibrium coils since the flow is too weak to suppress transverse fluctuations. The onset of intermolecular overlap  occurs for such stretched conformations at polymer concentrations much smaller than $\cstar$, the conventional critical overlap concentration for equilibrium coils.  Therefore, for a range of concentrations $c < \cstar$, intramolecular hydrodynamic interactions may be significantly screened in stretched conformations. Scaling arguments based on ``blob" concepts are used here to argue that the stretched state  drag coefficient can grow strongly with concentration in the dilute regime. A dumbbell model with conformation-dependent drag model is used to predict a concomitant strong  enhancement of coil-stretch hysteresis with increasing concentration  in the dilute regime. This extensional flow induced self-concentration leads to a maximum in hysteretic effects around $\cstar$, which progressively diminish in the semi-dilute regime where screening in isotropic coils reduces the difference in drag coefficient between stretched and coiled states.  It is shown that the concentration dependence observed by \citet{clasenetal} of capillary-thinning dynamics in liquid bridges of polymer solutions provides direct evidence of coil-stretch hysteresis enhancement by self-concentration. 
\end{abstract}

\maketitle
\section{\label{s:intro} Introduction}
Considerable theoretical insight has been achieved in recent years in understanding the dynamics of \textit{isolated} flexible homopolymers in strong flows \citep{larsonreview}. It is recognized that besides chain connectivity and entropic resistance to stretching, solvent-mediated intramolecular hydrodynamic interactions play a dominant role in determining polymer dynamics even in strong flows where polymer molecules are significantly stretched.  Nevertheless, our understanding of the physics at the level of \textit{isolated} molecules has not yet translated into an ability to fully and accurately predict rheological behaviour of \textit{dilute} polymer solutions; although much is known about their linear viscoelastic behaviour, systematic quantitative agreement of predictions from theory and simulations with experimental observations has been relatively rare well outside the equilibrium state \citep{larsonreview}.  For instance, experiments in recent years have taken advantage of the predominantly extensional flow generated in slender filaments thinning under capillary action to probe dilute solution rheology in strongly stretching flows \citep{mckinleysridhar}. The mismatch between experiment and theory  has been dramatically brought to the fore by \citet{clasenetal} in experiments studying dynamics of capillary-thinning in liquid-bridges of dilute polymer solutions.

In such experiments, a liquid-bridge is created from a drop of a test fluid sandwiched between a  pair of end-plates, by rapidly separating out the plates to a fixed distance \citep{bazilevsky1, bazilevsky2, liang, annamckinley}. If this distance is large enough, the bridge  undergoes capillary-thinning due to the Rayleigh-Plateau instability.  Polymer molecules caught in  the extensional flow that develops at the necking plane undergo the coil--stretch transition. Large elastic stresses thus develop, resisting capillary action and considerably slowing down the rate at which the bridge thins.  

\citet{entovhinch} used a stress-balance between capillary, viscous and elastic stresses to analyze inertialess capillary-thinning in a viscoelastic liquid. Using a viscoelastic model with a fixed relaxation spectrum, they showed that with exponential transient growth of the polymeric elastic stress (due to strain-hardening (SH)) becomes significantly larger than the viscous stress from the Newtonian solvent, the filament radius $R$ decays exponentially with time $t$, \textit{i.e.} $R \sim e^{-t/3 \lamo}$, where $\lamo$ is the longest relaxation time; in contrast, a viscous Newtonian liquid bridge thins linearly with time. The strain-rate at the necking plane for a slender filament is related to radial kinematics as 
\begin{gather}
\edot = - \frac{2}{R} \, \frac{d R}{d t}\,.
\label{e:strainrate}
\end{gather}
Further, the Weissenberg number for any extensional flow of strain-rate  $\edot$ is in general \textit{defined} as
\begin{gather}
\Wi = \edot \lamo \,.
\label{e:Widef}
\end{gather}
The Entov--Hinch prediction of exponential decay of filament radius thus implies that the strain-rate and Weissenberg number in the elastic regime during capillary-thinning are $\edote = 2/( 3 \lambda_0)$, and  $\Wie  =  2/3$, respectively.

Early experiments clearly confirmed the existence of an exponential-decay regime in capillary thinning of liquid bridges of polymer solutions \citep{liang, annamckinley}. However, \citet{clasenetal}  also observed with solutions of polystyrene in high-viscosity solvents that necking in the elastic regime proceeded such that $R \sim e^{-t/3 \lame}$, with a relaxation-time $\lame$ quite different from the $\lamo$ characterized using small-amplitude oscillatory shear (SAOS) experiment. This implies that the observed strain-rate in the elastic regime in these experiments, $\edote = 2/( 3 \lame)$ corresponds to a Weissenberg number (Eqn.~\eqref{e:Widef}), 
\begin{gather}
\Wie = \frac{2/3}{\lame/\lamo} \,,
\label{e:Wie}
\end{gather}
that is different from the value of $2/3$. These measurements were made across a wide range of polymer concentrations well in the dilute regime, and for several different polymer molecular weights. Similar observations have been reported on slender filaments of low-viscosity (inertia-dominant) dilute polyethylene-oxide (PEO)-in-water solutions \citep{viyaada}.  Classical constitutive models for  polymer stresses in dilute solutions such as the Oldroyd-B or FENE-P models---Entov and Hinch used a variant in their study---assume that the relaxation spectrum is solely dependent on molecular weight, but is otherwise fixed, which is contradicted by experimental observations in capillary-thinning that $\lame \neq \lamo$.  The trouble is that, once the relaxation spectrum is determined from SAOS measurements, these dilute solution models intrinsically have no mechanism that permit the relaxation time or spectrum to vary within a single sample.

An even more radical challenge to such models comes from the concentration dependence of $\lame$ observed in these experiments. In solutions with polymer concentrations $c \ll \cstar$, the critical overlap concentration, $\lamo$ is observed to depend only weakly on concentration,  as expected for solutions in the dilute regime, asymptotically approaching the value $\lamZ$ predicted by the Zimm theory for hydrodynamics of single polymers \citep{doiedw, dpl2} in the dilute limit ($c \rightarrow 0$). In sharp contrast, $\lame$ increases almost linearly with $c$ over the same range of concentrations. A dilute solution is, by definition, one where molecules are well separated from each other so that intermolecular interactions have negligible impact. The relaxation spectrum and hence $\lamo$ in such a solution are expected to be determined completely by average single molecule behaviour, and thus be independent of concentration. Conversely, a relaxation-time that is  strongly concentration-dependent suggests the presence of strong intermolecular interactions. The observation that $\lame$ is concentration-dependent while $\lamo$ for the same samples is not, indicates that intermolecular interactions increase substantially within a single sample in the extensional flow during capillary thinning: polymer solutions ``self-concentrate"! 

This challenges the current modeling paradigm of classifying polymer solutions into distinct concentration regimes based on $\cstar$ determined from equilibrium size of polymeric coils, and then developing models for non-linear rheology in those regimes. \citeauthor{clasenetal} hypothesized that when polymer molecules stretch significantly, they interact hydrodynamically with each other more strongly than equilibrium  coils at the same polymer concentration.  This suggestion was also motivated by observations by \citet{stoltz} of increased inter-chain interaction in multi-chain  BD simulations of dilute polymer solutions in strong shear and extensional flows. The underlying reasons for the increased between stretched chains are however not yet well understood.

The key to explaining these intriguing observations may lie in another observation typical of  capillary thinning \citep{clasenetal, viyaada} that $\lame$ is considerably larger than $\lamo$ for a substantial range of concentrations in the dilute (in the conventional sense) regime. From Eqn.~\eqref{e:Wie}, this means that the observed values of $\Wie$ are below the Entov--Hinch prediction of 2/3. In fact, it can be shown (as will be later, in Fig.~\ref{f:lamebylamz} (a)) from the observed values of $\lame$ and $\lamo$ that $\Wie$ in these experiments are well below the critical value of $\Wics = 1/2$ for the coil--stretch transition. These low values of $\Wie$ are found to be sustained for long durations (and strains). As pointed out earlier, the observed exponential decay of necking radius also means significant polymer stretching and large polymer contribution to fluid stresses at the necking plane. The combination of sustained sub-critical $\Wie$ and large polymeric stresses in an extensional flow suggests a role for coil--stretch \textit{hysteresis}. 

The seminal work of De Gennes, Hinch and Tanner \citep{degennes, hinch, tanner} showed that conformation-dependence of polymeric friction coefficient leads to hysteretic behaviour in conformational and rheo-optical properties  within a window of extension rates. Steady state in extensional flow is primarily the result of a balance between internal entropic resistance of flexible polymer molecules to stretching  and the frictional drag force exerted on molecules by the flowing solvent. Within the hysteresis window, non-linearities in the dependence of the entropic resistance and drag forces on molecular stretch cause the balance between internal resistance and drag to occur at two distinct values of the stretch \citep{schroeder_science, schroeder, sridharPRL}. One of these stable states corresponds to weakly deformed coils,  while in the other, molecules are highly stretched. \citet{degennes} showed that a large non-equilibrium free-energy barrier separates these two stable states for long molecules. Hysteretic behaviour thus emerges as molecules are drawn to either of  these stable states depending on initial conditions \citep{degennes, schroeder_science, schroeder, sridharPRL, hsiehlarsoncsh}.  The coiled state becomes unstable when $\Wi = \Wics = 1/2$ and above, and molecules rapidly unravel and stretch out. The lower bound of the hysteresis window on the other hand corresponds to a critical strain rate below which the free-energy barrier vanishes and stretched molecules will  always quickly relax to the coiled state. The critical Weissenberg number for this stretch--coil transition is denoted here as $\Wisc$. First predicted in the 1970s, the existence of coil--stretch hysteresis was confirmed in single-molecule experiments \citep{schroeder_science} and BD simulations \citep{schroeder, hsiehlarson, prabhakar_gaussian,sridharPRL} in extensional flows.

Classical constitutive models that assume constant friction, and hence, a fixed relaxation spectrum, correctly predict a coil--stretch transition in extensional flows at $\Wics = 1/2$, but without any hysteresis. Conformation-dependent drag thus provides for a natural mechanism for changes in relaxation times with flow, and also leads to hysteresis and the possibility of sustaining stretched states at low $\Wi$ in extensional flows, thus pointing to a possible resolution to questions posed by observations in capillary thinning. \citet{prabhakar} therefore coupled the mid-filament stress-balance of Entov and Hinch with a multi-mode constitutive model for a dilute polymer solution that accounted for conformation-dependent \textit{intramolecular} hydrodynamic interactions in bead-spring chains.  This led to an important finding: $\Wie$ in the elastic regime of capillary-thinning is not limited by 2/3, but rather by the value of $\Wisc$, which is lower than  $\Wics = 1/2$. 

While this provided evidence that coil--stretch hysteresis allows $\Wie$ to be significantly smaller than $1/2$, \citet{prabhakar} could not predict the strong dependence of $\lame$ and $\Wie$ on concentration observed in experiments. But their dilute-solution model did not account for any \textit{inter}-chain interactions; with any such model, the width of the hysteresis window  depends solely on chain length, but is independent of concentration. Indeed, at infinite dilution, the ratio $\Wics/\Wisc$ is proportional to ratio of the mean friction coefficient of a fully-stretched rod of length $L$ (predicted well by Batchelor's \citep{batchelor1} for dilute suspensions of slender rods),  to  $\zetaZ$, the (Zimm) value for isolated, isotropic coils at equilibrium. Therefore, in the dilute limit (and for solutions close to the theta state), 
\begin{gather}
\frac{\Wics}{\Wisc} \sim \frac{\sqrt{\NK}}{ \ln \NK} \,,
\label{e:windowwidth}
\end{gather}
where $\NK$ is the number of Kuhn segments in a flexible molecule  \citep{schroeder_science, schroeder, sridharPRL}. With such a classical dilute-solution model with conformation-dependent drag therefore, and with $\Wie \sim \Wisc$ in capillary thinning,  Eqns.~\eqref{e:Wie} and \eqref{e:windowwidth} above show that for any given molecular weight,  the $\lame$ predicted mirrors the lack of significant concentration dependence of $\lamo$ in the dilute regime. 

To this author's best knowledge, the concentration dependence of the width of the coil--stretch hysteresis window has not been studied before. In a highly\textit{ concentrated} polymer solution, it is known that  interpenetration of molecules completely screens out solvent-mediated hydrodynamic interactions \citep{doiedw, rubinsteincolby}. Any single molecule thus follows Rouse dynamics as opposed to Zimm hydrodynamics of isolated chains in dilute solutions. Since the Rouse friction coefficient $\zetaR$ of a molecule is independent of conformation, hysteresis is expected to vanish in concentrated polymer solutions.  If stretched molecules indeed interact more strongly than coiled molecules as suggested by  \citet{clasenetal},  their friction coefficient could also depend more strongly on concentration. Recalling that coil--stretch hysteresis arises essentially because of differences between friction coefficients of molecules in  stretched and coiled conformations, the stretched-to-coiled friction coefficient ratio could depend on concentration in solutions that are conventionally dilute under quiescent conditions.  In turn, this should give rise to a concentration-dependence in $\Wisc$ at the SCT. Further, from the finding of \citet{prabhakar} that $\Wie = \Wisc$ in capillary-breakup, $\Wie$ and therefore $\lame$ would depend strongly on concentration, even when $\lamo$ does not, in the dilute regime.  

Testing the hypothesis outlined above against the data of \citeauthor{clasenetal} is the primary objective of the present work. But what is the origin of the increased interaction between stretched polymers which are well separated as coils under quiescent conditions? Section~\ref{s:model} below uses insights from  single-molecule BD simulations of polymer chains to propose a mechanism for self-concentration in dilute polymer solutions.  Scaling arguments for hydrodynamic screening---well-established for semi-dilute solutions of coiled molecules---are then adapted for stretched chains. These results are then combined with other known scaling results to develop an interpolation scheme for the frictional drag as a function of chain stretch and concentration. This conformation-dependent friction is used then in the FENE-P dumbbell model \citep{dpl2} for viscoelastic stress in polymer solutions. Predictions for the concentration-dependence of steady-state coil--stretch hysteresis in uniaxial extensional flow are discussed first in Section~\ref{s:CSH}. The new constitutive model is used then in conjunction with a mid-filament stress balance to predict capillary thinning of nominally dilute as well semi-dilute solutions; these results are discussed in Section~\ref{s:capthin}, and are compared with experimental data of \citet{clasenetal}.  Section~\ref{s:concl} summarizes the principal conclusions of this article.

\section{\label{s:model} The Conformation-Dependent Drag model}
If the fluctuating end-to-end vector of a polymer molecule is $\bQ$, then the average shape of a molecule in solution can be characterized by the second moment of the probability distribution for $\bQ$, $\Mtens = \langle \bQ \bQ \rangle$, the angled brackets representing an ensemble average. In the equations below,  $n$, $\kB$, and $T$ are the polymer number density,  the Boltzmann constant, and the absolute solution temperature, respectively; $L$ is the polymer contour length; $M = \tr\, \Mtens =\Mzz + 2 \Mrr = \langle \widetilde{Q}^2 \rangle$ is the mean-squared end-to-end distance in general, and the subscript `0' indicates the equilibrium state of a polymer solution at a given $n$. In principle, the equilibrium size, and hence $M_0$, are dependent on concentration, since in a good solvent, the equilibrium size is expected to decrease as concentration is increased beyond $\cstar$ and excluded-volume interactions are screened \citep{doiedw, rubinsteincolby}. In the model here, excluded-volume (EV) effects are ignored; EV screening is therefore not relevant in such systems, and $M_0$ is effectively independent of concentration. As such, predictions here are valid only for solutions close to the theta state.   For chains close to the theta state, the number of Kuhn segments and the Kuhn length are defined as follows, respectively:
\begin{gather}
\NK = \frac{L^2}{M_0} \,; \quad  \bK = \frac{M_0}{L} \,.
\label{e:NKbK}
\end{gather}
In the FENE-P dumbbell model \citep{dpl2}, it is customary to introduce the following two parameters in connection with the entropic resistance to molecular stretching: the ``spring" constant $H$ and the finite-extensibility (FE) parameter,
\begin{gather}
b = \frac{L^2 \,H}{\kBT}\,.
\end{gather}
Here, these two parameters have been eliminated in favour of $M_0$ and $L$; using the equilibrium prediction of the FENE-P model, we can obtain
\begin{gather}
H = \frac{3 \, \kBT}{M_0} \, \left(1 - \frac{M_0}{L^2} \right) =  \frac{3 \, \kBT}{M_0}\,\left(1 - \frac{1}{\NK} \right) \,,
\end{gather}
and further,
\begin{gather}
b = 3 \,(\frac{L^2}{M_0} - 1) =3 \,(\NK - 1) \,.
\end{gather}

Equations for the conventional FENE-P model governing the evolution of components of $\Mtens$ can be modified to allow for a conformation-dependent friction coefficient $\zeta$ that is different from $\zetac$,  the average drag coefficient at equilibrium and at a given polymer number density $n$ \citep{degennes, hinch, tanner, dunlapleal, phanthien, bird1985}. Defining the near-equilibrium relaxation time as
\begin{gather}
\lamo = \frac{\zetac\,M_0}{12\, \kBT} \,,
\label{e:lamodef}
\end{gather} 
the modified equations for the components $\Mzz$ and $\Mrr$ in a uniaxial extensional flow are:
\begin{align}
\frac{d \Mzz}{d t} &= 2 \,\edot\, \Mzz - \frac{1}{\lamo \, (\zeta/\zetac)}\, \left( \,f\, \Mzz - \frac{M_0}{3} \right) \,,\label{e:mxxode} \\[3mm] 
\frac{d \Mrr}{d t} &= - \edot\, \Mrr - \,\frac{1}{\lamo \, (\zeta/\zetac)}\, \left( \,f\, \Mrr  - \frac{M_0}{3} \right)\,,
\label{e:myyode}
\end{align}
where $z$ represents the axial coordinate along the direction of extension, and $r$ is the radial coordinate in the transverse direction. The ratio,
\begin{gather}
f(M) = \frac{L^2 -M_0}{L^2 - M}  \,,
\label{e:fdef}
\end{gather}
represents the effect of FE on stiffness of entropic resistance to stretching. At equilibrium ($\edot = 0$; $t \rightarrow \infty$), $\Mzz = \Mrr = M_0/3$ is a solution to the equations above, and $f = 1$. In an imposed extensional flow, $f$ diverges as $M \rightarrow L^2$, preventing $M$ from exceeding $L^2$ in the model.

In polymer solutions, the flow-induced stress tensor $\bm{\uptau} = -\etas \, \bm{\dot{\upgamma}} + \bm{\uptau}_\mathrm{p}$, where $\etas$ is the viscosity of the Newtonian solvent, and $\taupt$ is the polymer contribution to the extra stress.  The only part of  the stress tensor that is rheologically relevant in uniaxial extensional flows is the first normal-stress difference, $\taup{zz}  - \taup{rr}$. The Kramers' equation for polymer stress then gives \citep{dpl2}:
\begin{gather}
\NIp = \taup{zz} - \taup{rr} = - \frac{3 \,\nkBT }{M_0}\, f \,(\Mzz - \Mrr)\,;
\label{e:Kramers}
\end{gather}
$\NIp$ vanishes at equilibrium. The polymer contribution to the extensional viscosity at any strain rate $\edot$ is defined as:
\begin{gather}
\eetap = - \frac{\NIp}{\edot} \,.
\end{gather}
The transient extensional viscosity, denoted as $\eetapp$, is analogously defined when either just $\NIp$,  or both $\NIp$ and $\edot$ are time-dependent. The ODEs~\eqref{e:mxxode} and \eqref{e:myyode} can be integrated to steady-state to obtain predictions for steady uniaxial extensional flow at a fixed value of $\edot$. For obtaining predictions for  capillary-thinning of viscous liquid-bridges of polymer solutions, the ODEs above are coupled to the following stress-balance at the necking plane, obtained after neglecting the effects of fluid inertia, gravity and axial curvature of the bridge \citep{entovhinch}:
\begin{gather}
\frac{\gamma}{R} - 3 \etas \edot + \NIp = 0\,,
\label{e:EHbal}
\end{gather}
where $\gamma$ is the surface tension coefficient, and the strain-rate at the mid-filament is given by Eqn.~\eqref{e:strainrate}. 


The Kramers' expression for polymer stress is strictly valid only in the dilute limit. A formal extension of the theory for solutions where intermolecular interactions are important can be expected to lead to terms explicitly of higher order in $n$.  Instead, a mean-field approach is taken here which models only the average behaviour of a single molecule in an effective medium consisting of the rest of the solution \citep{kroger}.  Intermolecular interactions enter the model through their influence on $\lamo$ and the ratio $\zeta/ \zetac$. For a given polymer, the coil-state friction $\zetac$ and relaxation time $\lamo$ do not vary with the conformation and depend significantly on the polymer concentration only in the semi-dilute regime, whereas, as indicated in the Introduction, $\zeta$ is anticipated to depend both on conformation and concentration. The evolution of $\Mzz$ and $\Mrr$ is thus directly influenced by polymer concentration, which brings about a nonlinear dependence of predictions of $\NIp$ on $n$.  

Dumbbell models \citep{degennes, hinch,tanner,fullerleal,phanthien} with conformation-dependent friction are built on the observation that as chains stretch in flow, the mean friction coefficient $\zeta$ should increase monotonically from $\zetac$ for equilibrium coils towards the value for fully-stretched chains. Partially stretched chains are pictured as inter-penetrable ``rods" of length $\sqrt{\Mzz}$ and diameter $\sqrt{\Mrr}$; for notational ease, we introduce $\ell = \sqrt{\Mzz}$ and $d = \sqrt{\Mrr}$. Analysis of frictional properties of rods typically assume that they are slender, that is, $\ell \gg d$. Since this assumption may not always be valid for partially unravelled chains, a simple linear weighted average is used here to interpolate $\zeta$ for partially stretched chains between a value $\zetac $ for equilibrium coils, and an estimate $\zetar$ for slender rods of length $\ell$ and diameter $d$: 
\begin{gather}
\zeta  = \, \frac{ L - \ell}{L - d_0}  \, \zetac + \, \frac{\ell - d_0}{L - d_0}\, \zetar \,.
\label{e:mixrule}
\end{gather} 
The sections below discuss the modeling of $\zetac$ and $\zetar$ for various concentration regimes. The case of isolated chains is taken up first. 

\subsection{\label{s:infdil} Partially stretched chains at infinite dilution}
At equilibrium ($\ell = d = d_0 = \sqrt{M_0/3}$), the average friction coefficient of an isotropic coil  $\zetac = \zetaZ$ at infinite dilution. For rod-like conformations, $\zetar$ is approximated by Batchelor's \citep{batchelor2} expression for slender rods:
 \begin{gather}
\zetar \sim \frac{\etas \,\ell}{\ln (\ell/d)} \,.
\label{e:batchelor-dillim}
\end{gather}
The scaling result above for $\zetar$ cannot be used directly since the logarithmic term becomes singular when $\ell = d$. Since the Zimm friction coefficient $\zetaZ \sim \etas\, d_0$, the following regularization is used:
\begin{equation}
\zetar = \frac{K}{K +  \,\ln (\ell/d)} \, \left(\frac{\ell}{d_0}\right)\, \zetaZ\,,
\label{e:zetadilutebatch}
\end{equation}
where $K$ is an empirical constant.

Variants of the CDD model in the dilute limit such as the above are known to reproduce coil--stretch hysteresis observed in single-chain BD \citep{degennes, hinch, tanner, fullerleal, phanthien, schroeder, thesis}. Appendix \ref{a:stretch} briefly discusses an approximate solution for the stretched state in steady uniaxial extensional flows. It is shown that a stretched state solution is obtained for $\Wi >  \Wisc$, and that $\Wisc < 1/2$ for sufficiently long chains.  The constant $K$ also controls the size of the hysteresis window in the dilute limit. Its value is chosen here to obtain a window size comparable to that observed in single-chain BD simulations; this is presented later in Fig.~\ref{f:BDScsh}.

\subsection{\label{s:pervol} Average pervaded volume}
Before proceeding further with modeling $\zeta$ for non-dilute solutions, it is necessary to examine the notions of molecular overlap and diluteness. At equilibrium, the average volume pervaded by an isotropic coil is $V_0 = d_0^3$. Therefore, for any given $n$, the volume fraction pervaded by polymers at equilibrium is
\begin{gather}
\phio \,=\, n \,d_0^3 \,.
\label{e:phio}
\end{gather}
The critical-overlap concentration $\cstar$ is conventionally estimated as the concentration at which coils just begin to overlap \textit{at equilibrium}, or when $\phio = 1$, which corresponds to a critical number density $n^\ast = 1/V_0$. This implies that the ratio $c/\cstar = n/n^\ast= \phio$, and henceforth, $\phio$ will therefore be used interchangeably with $c/\cstar$. The average molecular pervaded volume outside equilibrium for anisotropic conformations is estimated in the conformation-tensor model as $V = \ell d^2 = \sqrt{\Mzz} \Mrr$. The instantaneous pervaded volume fraction is hence
\begin{gather}
\phi = n \,\ell \,d^2 = \phio\,\frac{\ell \, d^2}{d_0^3} \,.
\label{e:phi}
\end{gather}
 
The ratio $c/\cstar$ is often used to classify polymer solutions into different concentration regimes. Equation~\eqref{e:phi} however shows that it is possible that molecular overlap may change with conformation. Upon the imposition of a flow gradient, the volume fraction $\phi$ will remain unchanged from $\phio$ only if molecular coils deform affinely with the incompressible solvent. Such affine deformation is only expected at high values of $\Wi$ and when FE effects are not important. But, during the elastic regime of capillary-thinning or within the coil--stretch hysteresis window, neither of these conditions may be true: as pointed out in the Introduction, $\Wi \lesssim O(1)$, and chains are further significantly stretched and FE may be important.

\begin{figure}
\centerline{\resizebox{12.7cm}{!}{\includegraphics{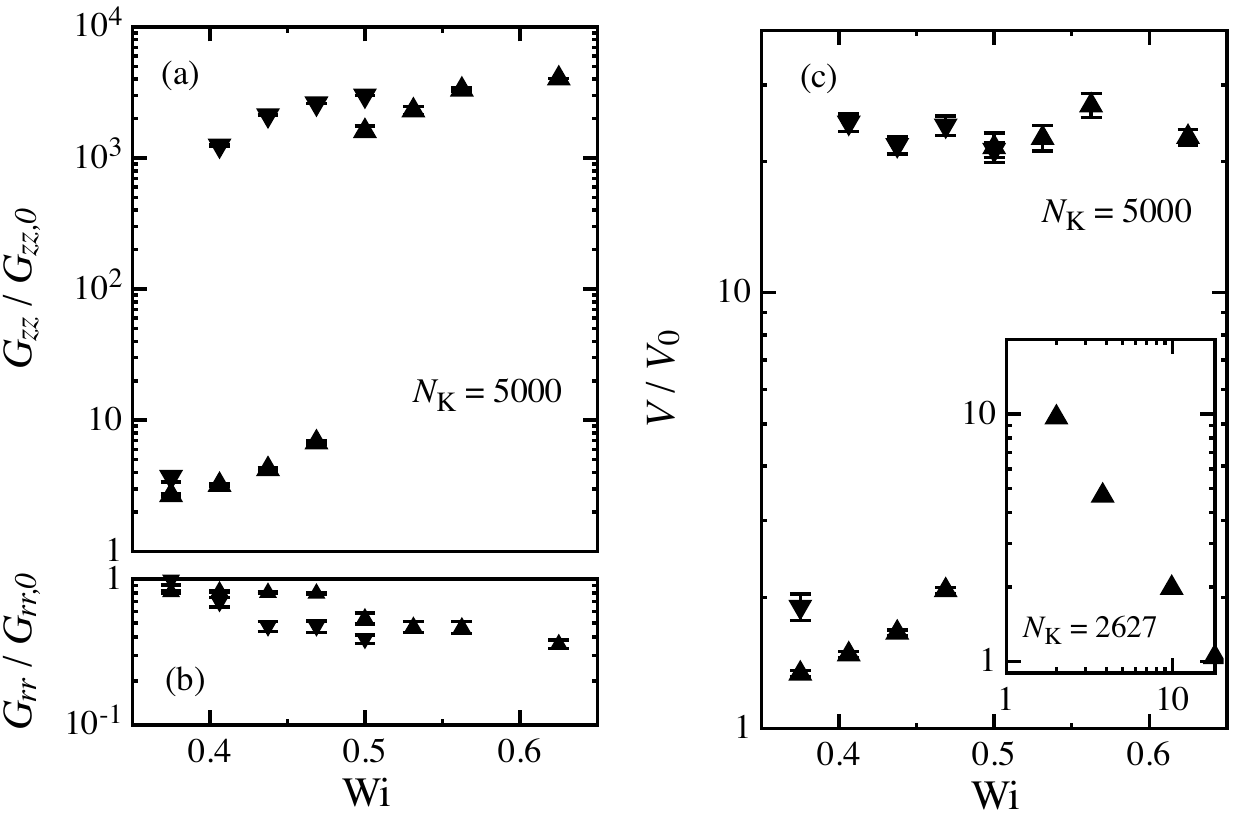}}}
\caption{\label{f:BDSgyration}  Predictions of coil--stretch hysteresis in (a) gyration tensor component  $G_{zz}$ and (b) $G_{rr}$, and (c) pervaded volume,  by BD simulations of isolated bead-spring chains: results are shown for simulations starting with initial ensembles  at equilibrium ($\filledmedtriangleup$), and with initially stretched to $90\%$ maximum permissible length ($\filledmedtriangledown$). These ensembles have different quasi-steady-states within the hysteresis window. Inset in (c) shows steady-state pervaded volume at high $\Wi$.} 
\end{figure}

Figure~\ref{f:BDSgyration} shows results of BD simulations in uniaxial extensional flow obtained by starting with ensembles of either equilibrium coils or highly stretched chains, and integrating to high strains at fixed values of $\Wi$.\footnote{The molecular model for these simulations and its parameters, and the simulation algorithm have been reported in Refs.~\citep{prabhakarmultiplicative, prabhakarSFG}. Briefly, a polymer molecule of $\NK$ Kuhn segments is modeled as a coarse-grained chain of beads connected by Finitely-Extensible Nonlinear Elastic (FENE) springs. Intramolecular hydrodynamic interactions are incorporated through Rotne-Prager-Yamakawa tensors. Excluded-volume interactions are neglected. Independent trajectories of chains are generated by integrating a stochastic differential equations for bead positions that describe their motion under the combined action of hydrodynamic drag forces exerted by a homogeneous flow of the surrounding solvent and its thermal fluctuations, and intramolecular connector forces.}  Coil--stretch hysteresis is clearly evident in the axial and radial components of the gyration tensor in Fig.~\ref{f:BDSgyration} (a) and (b) respectively, as well as in pervaded volume, plotted in Fig.~\ref{f:BDSgyration} (c).  

It is seen that in the stretched state (upright triangles; $G_{zz} \gg G_{zz,\,0}$) within the hysteresis window, the values of $G_{rr}$ are not much smaller than the equilibrium value $G_{rr,\,0}$.  This behaviour can be understood through the approximate solution of Eqns.~\eqref{e:mxxode} and \eqref{e:myyode}  presented in Appendix \ref{a:stretch}.  The transverse coil dimension in the stretched state $d_\mathrm{s}$  reflects the size of fluctuations in conformations transverse to the axis of the uniaxial extensional flow.  In the transverse direction, Brownian fluctuations tend to relax $d_\mathrm{s}$ towards the equilibrium value $d_0$. Opposing this, the motion of the solvent  drags chain segments radially inward towards the principal axis.  In addition, solvent motion along the principal axis tends to unravel and stretch a polymer chain. For stable stretched conformations, lengths are comparable to $L$, and FE results in increased chain stiffness, which also tends to suppress transverse excursions of a chain. From the analysis in the Appendix,
\begin{gather}
d_\mathrm{s}^2 = \frac{d_0^2}{\fs + \Wi \,(\zeta_\mathrm{s}/\zeta_0)} \,. 
\label{e:ds1}
\end{gather}
Since entropic resistance balances axial drag, $\fs$ is related to $\Wi$, and 
\begin{gather}
d_\mathrm{s}^2 = \frac{2\, \Wisc}{3\, \Wi\,\left[1 + \left(1 - (\Wisc/\Wi)\right)^{1/2} \right]}   \, d_0^2\,.
\label{e:ds2}
\end{gather}
Thus, for the stretched state, $d_\mathrm{s}$ is largest at the SCT where it is comparable to $d_0^2$, and decreases as $\Wi^{-1}$ at higher strain-rates. This indicates that in the vicinity of the SCT neither transverse solvent drag nor nonlinear stiffness effects are strong enough to significantly dampen transverse thermal fluctuations, explaining the observations in BD simulations (Fig.~\ref{f:BDSgyration} b). 

As an aside, Eqn.~\eqref{e:ds1} above is consistent with the observation that, for a chain stretched in the absence of flow by a tension applied to its ends, $d_\mathrm{s} \sim d_0$ as long as the effects of FE are not felt and $\fs \sim 1$ \citep{pincus1976, rubinsteincolby}. Further, Eqn.~\eqref{e:ds1} shows that the transverse size predicted by the ODEs \eqref{e:mxxode} and \eqref{e:myyode} is equivalent to a scaling estimate obtained by the blob argument that in strong flows transverse chain dimensions are the result of a balance between drag and transverse fluctuations \citep{colby2007}. We will return to this aspect later while calculating hydrodynamic screening lengths when stretched molecules overlap.

Large axial stretch and large transverse fluctuations  in the stretched state also imply that average volumes pervaded by these conformations are much larger than at equilibrium. Indeed, from Appendix~\ref{a:stretch}, the stretched-state pervaded volume,
\begin{gather}
V_\mathrm{s} = \sqrt{\frac{\NK}{3}}\,\frac{\Wisc}{\Wi}\, \, V_0 \,.
\label{e:Vs}
\end{gather}
It is largest at the SCT with $V_\Upsigma \sim V_0 \sqrt{\NK}$, and although it decreases as $\Wi^{-1}$ with extension-rate, at the coil--stretch transition $V_\mathrm{C} \sim  V_0 \,\ln \NK$ (since the ratio $\Wics/\Wisc \sim \sqrt{\NK}/\ln \NK$). The values of $\NK$ in the capillary-thinning experiments discussed in the Introduction are large with $\ln \NK \lesssim O(10)$ ($10^3 < \NK < 10^4$).  Hence, as observed in Fig.~\ref{f:BDSgyration} (c), stretched-state pervaded-volumes are substantially larger than those of equilibrium coils over a wide range of $\Wi$ including the hysteresis window. This also might explain observations by \citep{stoltz} in multi-chain BD simulations of stronger sensitivity to chain density in rheological properties around the coil--stretch transition than at low or high $\Wi$. They also observed more inter-chain crossings at $\Wi = 1$ than at $\Wi = 10$.  

The analysis above suggests that the pervaded volume fraction $\phi$ within the hysteresis window in an extensional flow can be significantly larger than $\phio = c/\cstar$ at equilibrium, and thus provides a simple mechanism for self-concentration. The calculation of $\zeta$ for non-dilute solutions must therefore account for the possibility of a dynamic crossover from one concentration regime to another during extensional flow. The following sections describe how  $\zeta$ is calculated given the parameters $n$, $M_0$ and $L$, and once the instantaneous chain dimensions $\ell$ and $d$ are known. 

\subsection{\label{s:coils} Friction coefficient of equilibrium coils}
The concentration dependence of $\zetac$ for non-overlapping equilibrium coils due to long-range intermolecular hydrodynamic interactions is not known exactly. In dilute limit (as $c \rightarrow 0$), it is expected that the polymer contribution to the zero-shear-rate viscosity $\etapo \sim n$, and the longest relaxation time $\lamo \sim \etapo/ (\nkBT)$ is thus independent of concentration. Therefore, from Eq.~\eqref{e:lamodef}, $\zetac$ is independent of the concentration and equal to $\zetaZ$ at low concentrations. As mentioned before, it is expected that $\zetac$ will not be very different from $\zetaZ$ when $0 < \phio < 1$. When $c > \cstar$, the phenomenon of hydrodynamic screening sets in.  Intramolecular hydrodynamic interactions arise due to the spread of velocity perturbations in the solvent created when any Kuhn segment moves. When chains interpenetrate, momentum in these velocity perturbations is dissipatively transferred to segments of neighbouring chains. Therefore, beyond a characteristic length-scale $\xih$ \citep{Gennes1976}, the presence of neighbouring chains screens out solvent velocity perturbations generated by one segment from reaching other segments of the same chain. The zone of hydrodynamic influence around each chain is thus pictured as that created by ``blobs" of size $\xih$ arrayed along its contour \citep{degennesbook, doiedw, rubinsteincolby}.

Although the existence of hydrodynamic screening is widely recognized based on observations in simulations and experiments, the underlying mechanism whereby it emerges in polymer solutions is still not fully understood \citep{ahlrichs2001}. A scaling estimate for the blob size $\xih$ can nevertheless be derived as the average size of the neighbourhood around any segment at which the number of segments from surrounding chains just begins to exceed that of its own chain. If the number of such blobs along any chain is $\Nxi$, then the number of Kuhn segments per blob is $\NK/\Nxi$. On the other hand, in a homogeneous solution, the mean segmental density is $n \NK$. Hence, the argument above for $\xih$ implies that
\begin{gather}
n \,\NK \, \xih^3 = \frac{\NK}{\Nxi} \,,\notag \\
\intertext{or,}
n \,\Nxi \,\xih^3 = 1 \,.
\label{e:Nxidef}
\end{gather}
Since $\Nxi \,\xih^3$ is the volume of the blobs on a single chain, the equation above means that blobs of all chains are space-filling. An additional independent equation for $\xih$ and $\Nxi$ comes from noting that, at equilibrium in a solution close to the theta state, chains obey ideal random-walk statistics, and
\begin{gather}
\xih^2 = \bK^2 \,\frac{\NK}{\Nxi} \,,\notag \\
\intertext{or,}
\Nxi \, \xih^2 = \bK^2 \,\NK = 3 \,d_0^2 \,,
\label{e:xihdef}
\end{gather}
 
Ignoring pre-factors,  Eqns.~\eqref{e:xihdef} and \eqref{e:Nxidef} can be solved to obtain $\xih$ and $\Nxi$ in terms of $\phio$ and $d_0$ (after substituting for $n$ in terms of $\phio$ using Eqn.~\eqref{e:phio}):
\begin{gather}
\xih = \frac{d_0}{\phio} \,; \quad \Nxi = \phio^2 \,, \text{ when } \phio \geq 1\,.
\label{e:xiNxi}
\end{gather}
Since intramolecular hydrodynamic interactions persist within each blob, the Zimm-like friction coefficient of each blob $\zetah$ is proportional to its size, and
\begin{gather}
\zetah = \left(\frac{\xih}{d_0}\right)\,\zetaZ  = \phio^{-1} \,\zetaZ \,.
\end{gather}
Further, hydrodynamic interactions are screened between blobs, and thus each chain behaves as a Rouse chain of $\Nxi$ blobs, and the overall average friction coefficient of an equilibrium coil is
\begin{gather}
\zetac = \zetah\, \Nxi = \phio\,\zetaZ \,.
\label{e:zetac1}
\end{gather} 
At critical overlap, each chain is contained within a single blob, whose friction coefficient is identical with the Zimm value for the whole chain. As $\phio$ increases further, the screening length $\xih$ decreases until it becomes comparable to a single Kuhn length, at which point screening of hydrodynamic interactions is considered to be complete. From Eqn.~\eqref{e:xiNxi}, the concentration at which $\xih = \bK$ and $\Nxi = \NK$ is
\begin{gather}
\phidagrc = \frac{d_0}{\bK} = \sqrt{ 3\, \NK} \,.
\label{e:phiodag}
\end{gather}
At this concentration, a polymer coil at equilibrium is fully Rouse-like, with a maximum friction coefficient $\zetaR$ obtained from Eqn.~\eqref{e:zetac1} as
\begin{gather}
\frac{\zetaR}{\zetaZ} = \phidagrc= \sqrt{3 \, \NK}\,.
\label{e:Rouse}
\end{gather}
For any $\phi > \phidagrc$, the friction coefficient is assumed to be constant, and equal to $\zetaR$.  

Putting these results together along with the negligible variation $\zetac$ below critical overlap, 
\begin{gather}
\frac{\zetac}{\zetaZ} \, = \, \begin{cases}
1  \,, \text{ if } \phio \leq 1 \,,\\
\phio \,, \text{ if } 1 < \phio \leq \phidagrc \,,\\
\sqrt{3\,\NK} \,, \text{ if } \phio > \phidagrc \,.
\end{cases}
\label{e:zetac2}
\end{gather}
Next, the model for $\zetar$ is described for its use along with $\zetac$ above in the weighted-average in Eqn.~\eqref{e:mixrule}. As mentioned earlier, $\zetar$ depends not only on the instantaneous conformational components $\ell$ and $d$, but also on $n$ through the instantaneous pervaded volume fraction $\phi$. Three separate concentration regimes are identified for calculating $\zetar$ depending on the value of $\phi$.  

\subsection{\label{s:batchelor1} Batchelor regime for partially stretched, non-overlapping chains}
Earlier, Eqn.~\eqref{e:zetadilutebatch} presented a regularization of  Batchelor's \citeyearpar{batchelor1, batchelor2} asymptotic result for slender rods (Eqn.~\eqref{e:batchelor-dillim}) at infinite dilution. When $\phio \neq 0$, the instantaneous volume fraction $\phi$ is calculated by Eqn.~\eqref{e:phi}, given $\phio$, $\ell$ and $d$. In a non-dilute suspension of rods of length $\ell$ and diameter $d$, all aligned along the extensional axis, the mean transverse separation $h$ between rods is such that $n \,\ell\, h^2 = 1$, and therefore from Eqn.~\eqref{e:phi},
\begin{equation}
\frac{h}{d} = \frac{1}{\sqrt{\phi}} \,.
\label{e:heqn}
\end{equation} 

Batchelor's result, Eqn.~\eqref{e:batchelor-dillim},  in the dilute regime is valid when $h \gg \ell$. In that case, inter-rod interactions are negligible. Weak hydrodynamic interactions between points along each rod lead to the logarithmic term in the denominator. For non-dilute rod suspensions, when $d \ll h \ll \ell$, \citet{batchelor2}  showed that the only change is that the logarithmic correction has $h/d$ as its argument: that is,
 \begin{gather}
\zetar \sim \frac{\etas \,\ell}{\ln (h/d)} \,.
\label{e:batchelor-nondillim}
\end{gather}
Effectively, $h$ acts as a screening length, and points along a rod farther apart than $h$ cannot ``see" each other hydrodynamically. \citet{batchelor2} suggested an interpolation between these results for dilute and non-dilute suspensions of the  following form, valid for all $h \gg d$ (or $\sqrt{\phi} \ll 1$):
\begin{gather}
\zetar \sim \frac{\etas \,\ell}{\ln (\ell/d) - \ln (1 + \ell/h)} \,.
\label{e:batchelor-nondil}
\end{gather}
This expression leads to an unphysical $\zetar < 0$ when $\ell/d \leq  1 + \ell/h$, that is, when $\ell/d \leq (1 - \sqrt{\phi})^{-1}$. Therefore, the following regularization similar to the one in Eqn.~\eqref{e:zetadilutebatch} is used instead:
\begin{equation}
\frac{\zetar}{\zetaZ}  = \frac{K}{K +  \,\ln G (\ell/d, \phi)} \, \left(\frac{\ell}{d_0}\right)\,,
\label{e:zetanondilbatch}
\end{equation}
where $K$ is the same constant as in Eqn.~\eqref{e:zetadilutebatch}, and the argument of the logarithmic correction in the denominator,
\begin{align}
G(\ell/d, \phi) \,=\, \begin{cases}
\,\displaystyle{\frac{\ell/d}{1 + \ell/h} \, = \, \frac{\ell/d}{1 + (\ell/d)\, \sqrt{\phi}}}\,, &\text{if }  \phi\,<\,(1 - d/\ell)^2\\[5mm]
\,1 \,.& \text{if } \phi \,\geq \,(1 - d/\ell)^2\\
\end{cases}
\label{e:Gcases}
\end{align}
This choice is consistent with Eqn.~\eqref{e:zetadilutebatch} when $n \rightarrow 0$, at a given $\ell$ and $d$. The equations above for $\zetar$ are used when $d/h = \sqrt{\phi}$ is small. The maximum value of $d/h$ for this Batchelor regime is chosen here to be 0.1; that is, Eqns.~\eqref{e:zetanondilbatch} and \eqref{e:Gcases} above are valid for $\phi \leq \phi_\mathrm{B} = 0.01$. With this choice, only the first case of Eqn.~\eqref{e:Gcases} is required for $G$ for all  $\ell/d > 1/0.9 = 1.11$ in the Batchelor regime.

\subsubsection{Transition from Batchelor regime to critical overlap}
The range $\phi_\mathrm{B} < \phi \leq 1$ then represents a transition between the Batchelor regime and the onset of critical overlap in stretched chains.  To suggest an expression for $\zetar$ in this regime, we consider the case when $\phi = 1$ and the solution consists of closely packed anisotropic ``rods" just in contact with their neighbours. For real rods, Eqn.~\eqref{e:batchelor-nondillim} with $h = 0$ is invalid; instead, frictional drag is dominated by lubrication interactions. A polymer coil however does not have a well-defined surface where the solvent must satisfy the no-slip condition in the same sense that a rod does, and lubrication effects between coils is possibly unimportant. Instead, it is assumed that when $h \lesssim d$, screening of the weak intermolecular hydrodynamic interactions along the length of each rods (which lead to the logarithmic correction term)  takes place and is complete when $h = 0$, so that when the chains just begin to overlap, each partially stretched chain behaves as a Rouse array of $\Nd = \ell/d$ ``beads" of size $d$ with no hydrodynamic interactions between these beads. Intramolecular interactions are restricted within each bead whose drag coefficient is proportional to its size, that is, 
\begin{gather}
\zetad = (d/d_0) \, \zetaZ\,.
\end{gather}
Therefore, at $\phi = 1$, the total rod-like friction coefficient is the total Rouse drag of $\Nd$ beads of friction $\zetad$:
\begin{gather}
\frac{\zetar}{\zetaZ} = \Nd\, \frac{d}{d_0} = \frac{\ell}{d_0} \,. 
\label{e:zetaovlp}
\end{gather}
For any $\ell$ and $d$, if the pervaded volume fraction is in the transition regime, that is, $\phi_\mathrm{B} < \phi \leq 1$, $\zetar$ is calculated by linearly interpolating with respect to $\phi$ between the values calculated by Eqn.~\eqref{e:zetasemidil} with $\phi = \phi_\mathrm{B}$, and by Eqn.~\eqref{e:zetaovlp} at $\phi = 1$:
\begin{align}
\frac{\zetar}{\zetaZ} &=  \left(\,\frac{1 - \phi}{1 - \phi_\mathrm{B}}\,\right)\,\left(\,\frac{\zetar}{\zetaZ}\,\right)_{\mathrm{B}}  \,+\,\left(\,\frac{\phi - \phi_\mathrm{B}}{1 - \phi_\mathrm{B}} \,\right)\,\frac{\ell}{d_0}  \,  \,, \notag \\
&= \left[ \, \left(\,\frac{1 - \phi}{1 - \phi_\mathrm{B}}\,\right)\,\left(\,\frac{K}{K + \,\ln \,G_\mathrm{B} (\ell/d)} \, \right) \,+\, \frac{\phi - \phi_\mathrm{B}}{1 - \phi_\mathrm{B}} \,\right] \, \frac{\ell}{\,d_0} 
\label{e:zetatrans}
\end{align}
where the subscript $\mathrm{B}$ indicates an evaluation with Eqns.~\eqref{e:batchelor-nondil} and \eqref{e:Gcases} for the Batchelor regime, at $\phi = \phi_\mathrm{B}$ and with the given $\ell$ and $d$.

\subsection{\label{s:batchelor2} Partially stretched, overlapping chains}
\begin{figure}[h!]
\centerline{\resizebox{12.7cm}{!}{\includegraphics{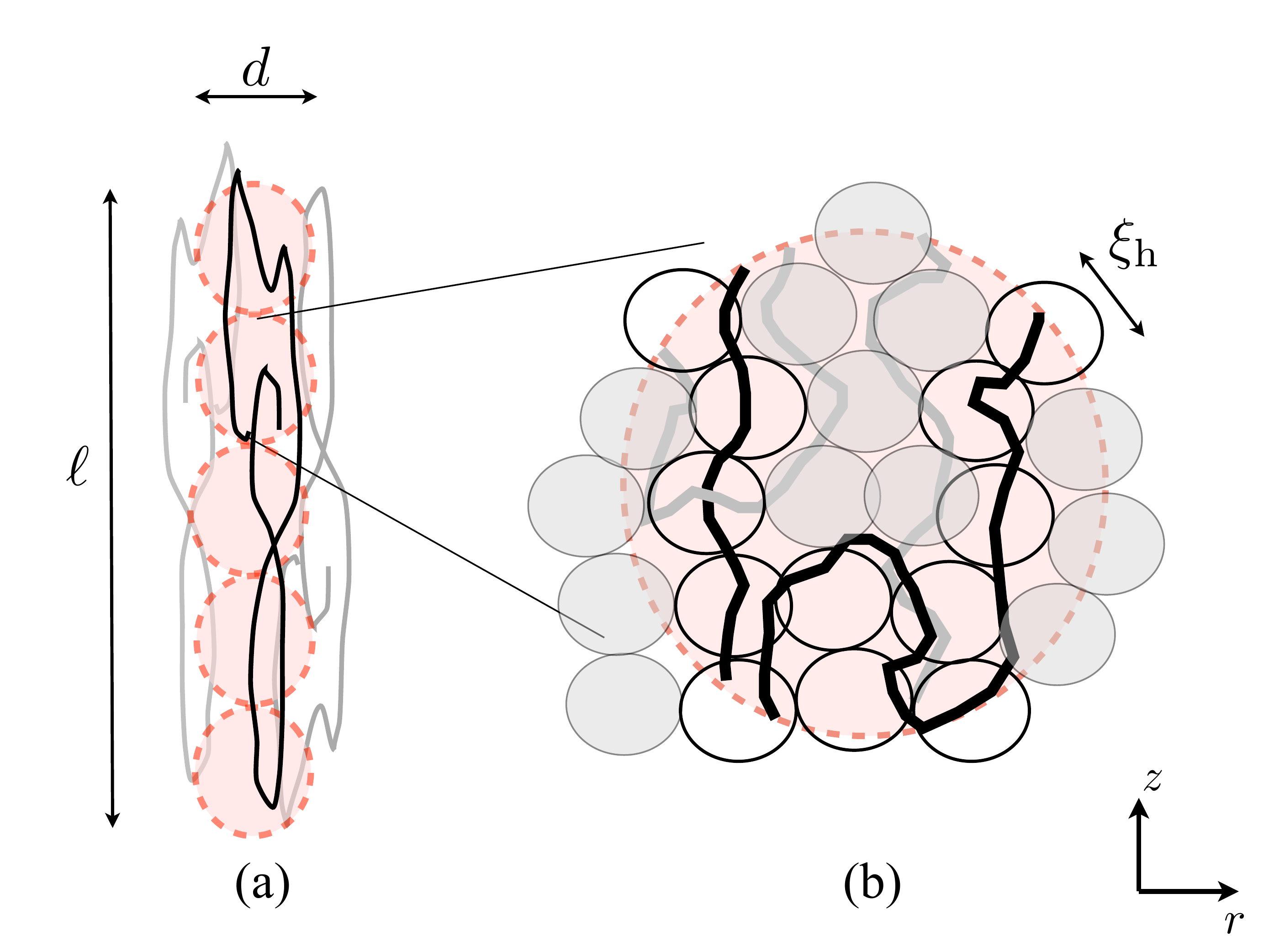}}}
\caption{\label{f:schematic} Schematic of length-scales in a solution of partially unravelled, overlapping polymer chains aligned along the principal stretching direction in a uniaxial extensional flow: the black curve in (a) and (b) indicates a test chain, while the grey curves are neighbouring chains. The partially unravelled chain in (a) is pictured as a linear array of beads, each the same size as that of conformational fluctuations $d$ transverse to the extensional axis. Within each bead, the test molecule is a chain of correlation blobs of size $\xih$. Hydrodynamic blobs from all chains are space-filling. In (b), blobs of the test chain are shown as white circles, while those of other chains are grey circles.}
\end{figure}

To my best knowledge, hydrodynamic screening within anisotropic chains when $\phi > 1$  has not been theoretically analyzed thus far. To do so, a partially stretched chain is still pictured as a linear array beads of size $d$ as mentioned above [Fig.~\ref{f:schematic} (a)]. As in the analysis of overlapping isotropic coils, each chain is further divided into $\Nxi$ blobs of size $\xih$ such that the same-chain segmental density within a blob is equal to the average segmental density across the whole solution; that is, hydrodynamic blobs of all chains in the polymer solution are space-filling as before, and Eqn.~\eqref{e:xihdef} involving $\xih$ and $\Nxi$ is still valid.

Further, recalling the discussion earlier of the stretched state, the transverse size $d$ is the length scale at which lateral thermal diffusion is balanced by the \textit{combined} effect of chain stiffness and solvent drag; $d$ thus represents a blob arising from the combination of nonlinear-stiffness and flow. It is reasonable to assume that within each bead of size $d$, lateral diffusion wins, and that the segmental distribution is equilibrium-like. Since $\NK/\Nxi$ and $\NK/\Nd$ are the numbers of Kuhn segments per blob, and per bead, respectively,
\begin{gather}
\xih^2 \, \sim \, \frac{\NK}{\Nxi}\,; \quad d^2\,\sim  \,\frac{\NK}{\Nd}\,.
\end{gather}
The difference between the assumption above for $\xih$ and the one leading to Eqn.~\eqref{e:xihdef} for equilibrium coils is that that here the pre-factor is not assumed to be $\bK^2$ but the square of another local length scale arising directional persistence due to axial alignment and stretching. Indeed, using $\bK^2$ on the right-hand side of the equation above for $d^2$ suggests that $\ell \, d $ is constant, which is inconsistent with the evolution equations for the conformation tensor components. 

With the assumption above, 
\begin{gather}
\frac{\xih^2}{d^2} = \frac{\Nd}{\Nxi} = \frac{\ell/d }{\Nxi}\,.
\label{e:blob2}
\end{gather}
Solving Eqn.~\eqref{e:blob2} and \eqref{e:Nxidef} for $\xih$ and $\Nxi$ gives for partially stretched chains,
\begin{gather}
\xih = \frac{d}{\phi} = \frac{d_0^3}{\phio\, \ell \, d} \,; \quad \Nxi = \phi^2 \,\frac{\ell}{d} = \phio^2 \,\frac{\ell^2 \, d}{d_0^3} \,, \text{ when } \phi \geq 1\,.
\label{e:xiNxi-2}
\end{gather}
These are consistent with Eqns.~\eqref{e:xiNxi} for equilibrium coils, recovering those results as $\ell \rightarrow d \rightarrow d_0$. At critical overlap, when  $\phi = 1$, $\xih = d$: each hydrodynamic blob is the same size as a bead.

As in the case of $\zetac$ for overlapping coils at equilibrium, $\zetar$ is calculated as the Rouse drag of $\Nxi$ blobs, each with a Zimm friction $\zetah = (\xih/d_0) \,\zetaZ$; hence,
\begin{equation}
\frac{\zetar}{\zetaZ} = \frac{\xih}{d_0} \Nxi = \phi \, \frac{\ell}{d_0}  = \phio \, \frac{\ell^2\, d^2}{d_0^4} \,.
\label{e:zetasemidil}
\end{equation}
For any given $\ell$ and $d$, $\zetar$ grows as $\phio$ with increasing chain overlap. This is not allowed to continue indefinitely, but is limited to the maximum value of the overall free-draining Rouse drag of a chain, given by Eqn.~\eqref{e:Rouse}. 

\subsection{\label{s:dimless} Dimensionless equations for the CDD model}
For steady extensional flows at a constant $\Wi = \edot\, \lamo$, there is no coupling of the conformational ODEs to an equation governing the macroscopic flow. The coupling of these equations above with the mid-filament stress balance in the case of capillary thinning is discussed later in Section~\ref{s:capthin}.  Rescaling the conformation tensor with $M_0$ as the characteristic conformation scale, and noting that $d \varepsilon = \edot \, dt$, the dimensionless ODEs for conformation components are:
\begin{align}
\frac{d \Mzz}{d \varepsilon} & = 2 \,\Mzz - \frac{1}{\Wi \, (\zeta/\zetac)}\, \left( \,f\, \Mzz - \frac{1}{3} \right) \,, \label{e:mxxode-d} \\[3mm]
\frac{d \Mrr}{d \varepsilon} & = - \Mrr - \,\frac{1}{\Wi \, (\zeta/\zetac)}\, \left( \,f\, \Mrr  - \frac{1}{3} \right)\,,
\label{e:myyode-d}
\end{align}
where,
\begin{gather}
f(M) = \frac{\NK -1}{\NK - M}  \,.
\label{e:fdef-d}
\end{gather}
In the absence of a macroscopic flow equation, the equilibrium relaxation time $\lamo$ at any given concentration $\phio$ is chosen as the characteristic time-scale, rather than its value $\lamZ$ at infinite dilution. With this choice, the coil--stretch transition always occurs at a fixed value of $\Wi = 1/2$, whether the solution is in the dilute or semi-dilute regimes. The ODEs above for $\Mzz$ and $\Mrr$ can be integrated to steady-state.  At any instant during the integration, the dimensionless $\ell$ and $d$ are known from the conformation-tensor components. Hence, the instantaneous $\phi$ is determined using Eqn.~\eqref{e:phi}. The dimensionless equivalent of the weighted-averaging in Eqn.~\eqref{e:mixrule} for the drag coefficient is written as:
\begin{gather}
\frac{\zeta}{\zetac}  = \, \frac{ \NK - \ell}{\NK - d_0}   + \, \frac{\ell - d_0}{\NK - d_0}\, \frac{\zetar}{\zetac} \,,
\label{e:mixrule-d}
\end{gather} 
where the dimensionless $d_0 = 1/\sqrt{3}$.The ratio $\zetar/\zetac$ is then evaluated by combining the expression for  $\zetac/\zetaZ$ given by Eqn.~\eqref{e:zetac1} and those for $\zetar/\zetaZ$, which are summarized in Table~\ref{t:drageqns}). Both these ratios are limited to a maximum of $\zetaR/\zetaZ$. For the case of steady extensional flow, $\nkBT$ is the characteristic stress scale. Thus, the rescaled polymeric first normal-stress difference
\begin{gather}
\NIp = \taup{zz} - \taup{rr} = - 3 \, f \,(\Mzz - \Mrr)\,.
\label{e:Kramers-d}
\end{gather}
The polymer contribution to the viscosity is rescaled by $\nkBT\, \lamo$ so that the dimensionless extensional viscosity is,
\begin{gather}
\eetap = - \frac{\NIp}{\Wi} \,.
\end{gather}
Predictions for the standard FENE-P model are obtained by setting $\zeta/\zetac = 1$ in the conformational ODEs. Thus, for steady flows, $\Wi$ and $\NK$ are the only model parameters in the FENE-P model, whereas the CDD model additionally requires $\phio = c/\cstar$. The multiple steady-states for coil--stretch hysteresis are generated by fixing $\Wi$ and using $\Mzz = \Mrr = 1/3$ as the initial conditions for predictions for the coiled state, and $\Mzz = 0.9 \NK$ and $\Mrr = 1/3$  as the initial conditions for the stretched state.  

\begin{table}[h]
\caption{\label{t:drageqns} Summary of equations for the ratio $\zetar/\zetaZ$}
\begin{center}
\begin{tabular*}{0.6\textwidth}{@{\extracolsep{\fill} }  c c}
\hline
\hline
 $\phi$             & $\zetar/\zetaZ$ \\
\hline\noalign{\smallskip}
$\phi \leq \phi_\mathrm{B}$ &$\displaystyle{\left[ \frac{K}{K +  \,\ln G} \right] \, \frac{\ell}{d_0} } $ \\[3mm]
$\phi_\mathrm{B} < \phi \leq 1$ &$\displaystyle{\left[ \frac{K}{K + \,\ln G_\mathrm{B}} \, \frac{1 - \phi}{1 - \phi_\mathrm{B}} \,+\, \frac{\phi - \phi_\mathrm{B}}{1 - \phi_\mathrm{B}} \right] \, \frac{\ell}{\,d_0}}$        \\[3mm]
 $\phi > 1$  &  $\displaystyle{\phi \, \frac{\ell}{d_0}}$\\[3mm]
\noalign{\smallskip}
\hline
\hline
\end{tabular*}
\end{center}
\end{table}

\section{\label{s:resdisc} Results and discussion}
Predictions for the concentration-dependence of steady-state coil--stretch hysteresis are presented first, followed by results for capillary-thinning. These are obtained for several values of $\phio = c/\cstar$ in the range $10^{-4}$ to 100.  All calculations are performed for a fixed value of $\NK = 5000$.

\begin{figure}[t]
\centerline{\resizebox{8.3cm}{!}{\includegraphics{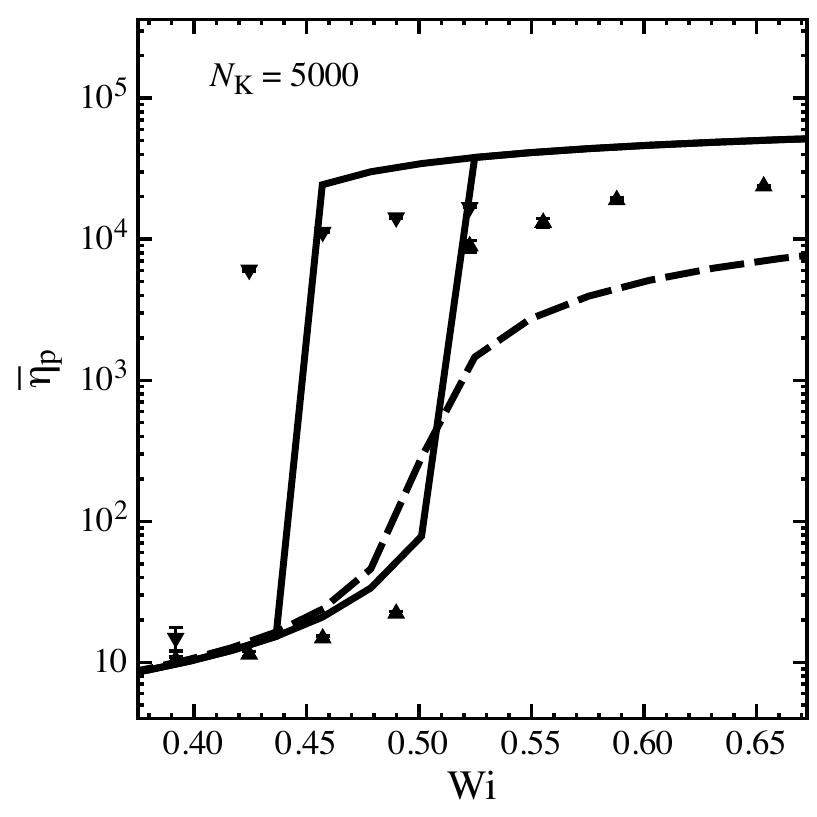}}}
\caption{\label{f:BDScsh}  Comparison of predictions of the CDD model (continuous curves) at infinite dilution for coil-stretch hysteresis in dimensionless extensional viscosity  with results of single-chain BD simulations: symbols are simulation results obtained with initial ensembles consisting of equilibrium coils ($\filledmedtriangleup$), and of initially stretched chains ($\filledmedtriangledown$); the FENE-P model (dashed curve) predicts no hysteresis.} 
\end{figure}

As the analysis in Appendix~\ref{a:stretch} shows, the size of the hysteresis window predicted by the CDD model depends on the constant $K$ in Eqn.~\eqref{e:zetanondilbatch} for the friction coefficient in non-overlapping chains. Its value is chosen here so that the hysteresis loop predicted in the dilute limit ($\phio = 0$) is  comparable in size to that observed in single-chain BD simulations of flexible chains  with $\NK = 5000$  (Fig.~\ref{f:BDScsh}). A value of $0.15$ gives good agreement both in the horizontal width and vertical height of the hysteresis window for the dimensionless polymer extensional viscosity, and is retained for all other calculations with the CDD model.

\subsection{\label{s:CSH} Self-concentration and coil--stretch hysteresis in steady uniaxial extensional flows}

\begin{figure}
\centerline{\resizebox{12.7cm}{!}{\includegraphics{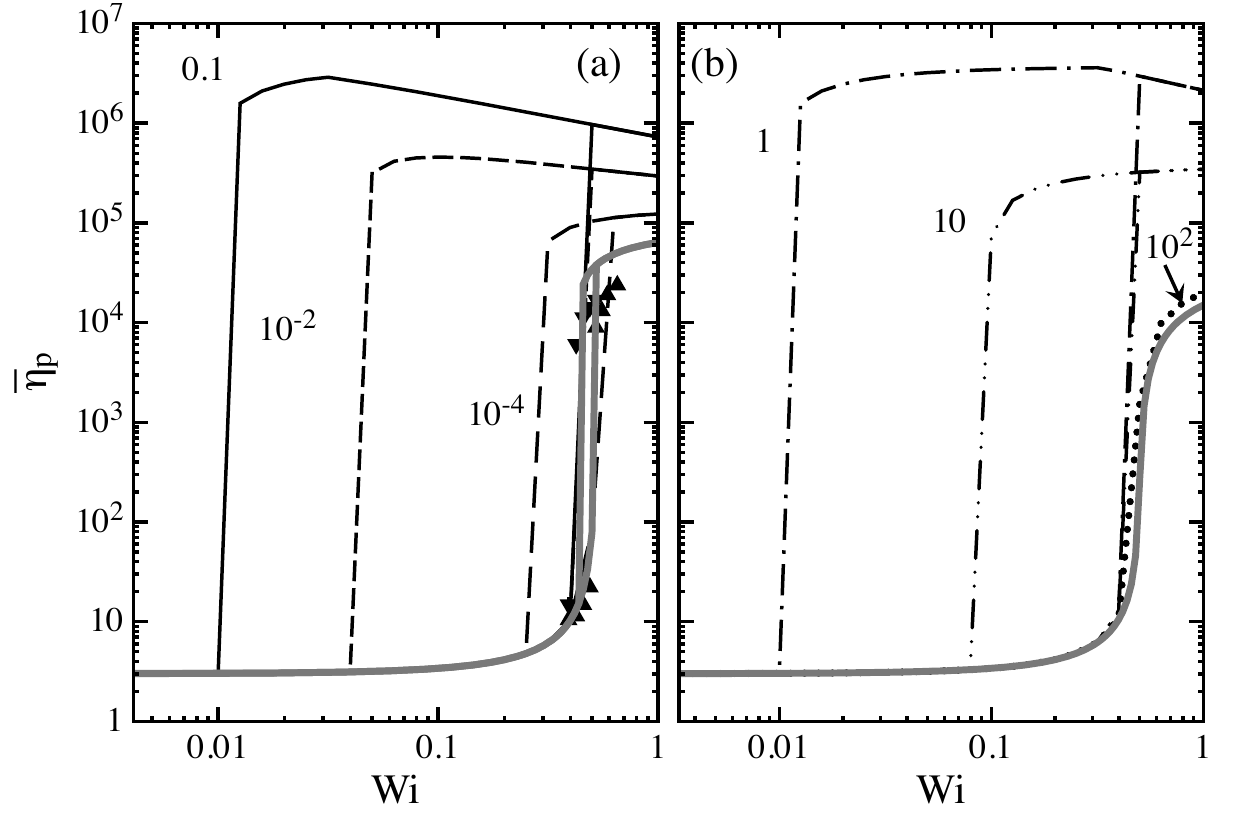}}}
\caption{\label{f:csh} Coil-stretch hysteresis in the dimensionless polymer contribution to steady-state extensional viscosity,  shown for the sake of clarity separately for (a) $\phio < 1$, and (b)  $\phio \geq 1$; numbers alongside curves indicate $\phio$ values. Filled symbols shown in (a) are results of single-chain (\textit{i.e.} for $\phio = 0)$ BD simulations.  The continuous grey curves in (a) are predictions obtained with the CDD model in the infinite-dilution limit, whereas the grey curve in (b) is the hysteresis-free prediction of the FENE-P model.}
\end{figure}

\begin{figure}[h!]
\centerline{\resizebox{8.3cm}{!}{\includegraphics{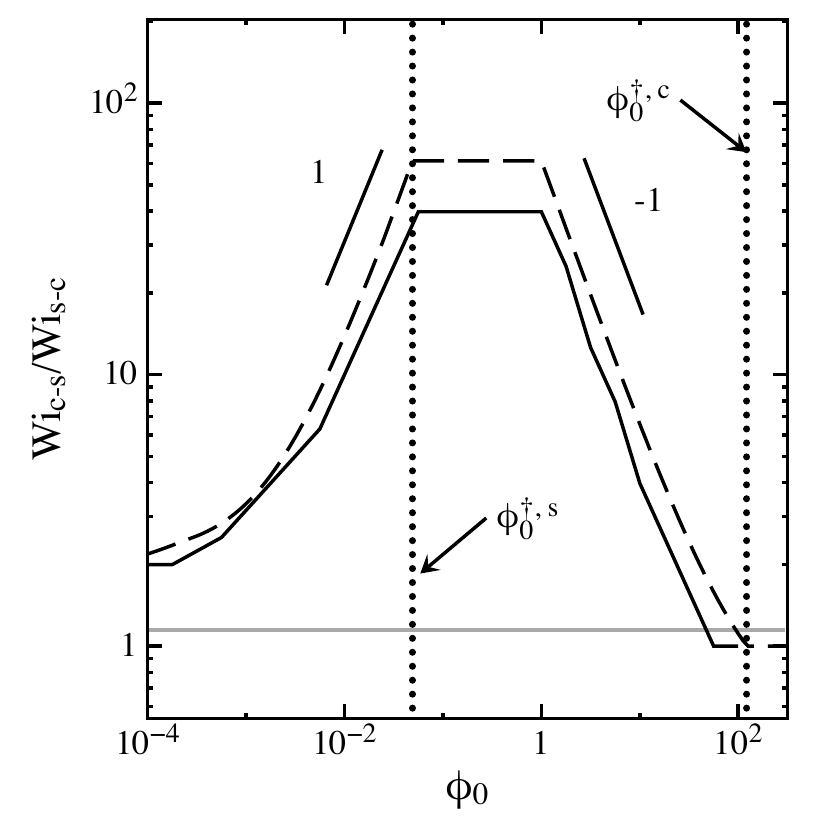}}}
\caption{\label{f:wiratio} Concentration-dependence of the width of the coil-stretch hysteresis window predicted by the CDD model (continuous curve): the horizontal line, $\Wisc/\Wics \approx  K \sqrt{\NK}/\ln \NK$ (= 1.25 for $\NK = 5000$ and $K = 0.15$), is the prediction at infinite dilution. The dashed curve is an upper bound estimated by calculating the ratio directly from the friction coefficient assuming that in the stretched state, the average axial stretch $\ells$ and transverse diameter $\ds$ are equal to estimates derived in Appendix~\ref{a:stretch} for  the SCT \textit{i.e.} $\ell_\Upsigma = L/2$ (Eqn.~\eqref{e:ellsc}), and  $ d_\Upsigma = \sqrt{2/3} \,d_0$ (Eqn.~\eqref{e:dsc}).}
\end{figure}

A strong concentration dependence is observed in the hysteresis windows predicted by the CDD model in Fig.~\ref{f:csh}.  For any choice of $\phio$, the coil--stretch transition always occurs at $\Wics \approx 0.5$, but $\Wisc$ for the SCT transition depends on polymer concentration. As mentioned above, the choice of $K$ ensures that the hysteresis window is similar to that observed in single-chain BD results in the limit of infinite dilution  (Fig.~\ref{f:BDScsh}; Fig.~\ref{f:csh} (a)). With increasing concentration,  $\Wisc$  first \textit{decreases} strongly  with increasing concentration in the range $10^{-4} < \phio < 0.1$ (Fig.~\ref{f:csh} (a)). In the range $0.1 < \phio < 1$, the width of the hysteresis window remains unchanged, but as $\phio$ increases above 1 (Fig.~\ref{f:csh} (b)), the hysteresis window begins to close as $\Wisc$ increases and approaches $\Wics \sim 1/2$. Hysteresis finally vanishes at large $\phio$ values, where predictions of the CDD and conventional FENE-P models become identical. The prediction of the width of the hysteresis window as quantified by the ratio $\Wics/\Wisc$ of critical Weissenberg numbers thus exhibits a broad maximum with respect to $\phio$ (continuous curve in Fig.~\ref{f:wiratio}), the plateau at maximum spanning almost a decade in concentration for $\NK = 5000$. The maximum hysteresis window size is much larger than that at infinite dilution (horizontal line in Fig.~\ref{f:wiratio}). 

As pointed out in the Introduction, it is known that from the analysis for dilute solutions that $\Wics/\Wisc$ is proportional to ratio of drag coefficients of stretched chains and equilibrium coils \citep{degennes, schroeder, thesis}. The analysis in Appendix~\ref{a:stretch} shows that 
\begin{gather}
\frac{\Wics}{\Wisc}\, =  \, \frac{\zeta_\Upsigma}{\zetac} \,,
\label{e:wiratio-zetaratio}
\end{gather}
where $\zeta_\Upsigma$ is the friction coefficient of the stable stretched state exactly at the SCT.  This specific configurational state is referred to henceforth as the $\Upsigma$-state. In this state, the axial and transverse chain dimensions are well approximated by (Eqn.~\eqref{e:ellsc} and \eqref{e:dsbyd0} respectively)
\begin{gather}
\ell_\Upsigma = \frac{L}{2} \,; \quad d_\Upsigma = \sqrt{\frac{2}{3}} \, d_0 \,.
\label{e:ellsc-dsbyd0}
\end{gather}
Since chain dimensions are known for the $\Upsigma$-state, it provides a handle on studying the scaling of the hysteresis-window size with respect to chain length and concentration.  Figure~\ref{f:wiratio} shows the estimate (dashed curve) for $\Wics/\Wisc$ directly obtained by using the chain dimensions above for calculating the friction coefficient $\zeta_\Upsigma$ as a function of $\phio$; this estimate compares well with the prediction obtained by integrating the ODEs of the CDD model to steady state (continuous curve). 

\begin{figure}
\centerline{\resizebox{12.7cm}{!}{\includegraphics{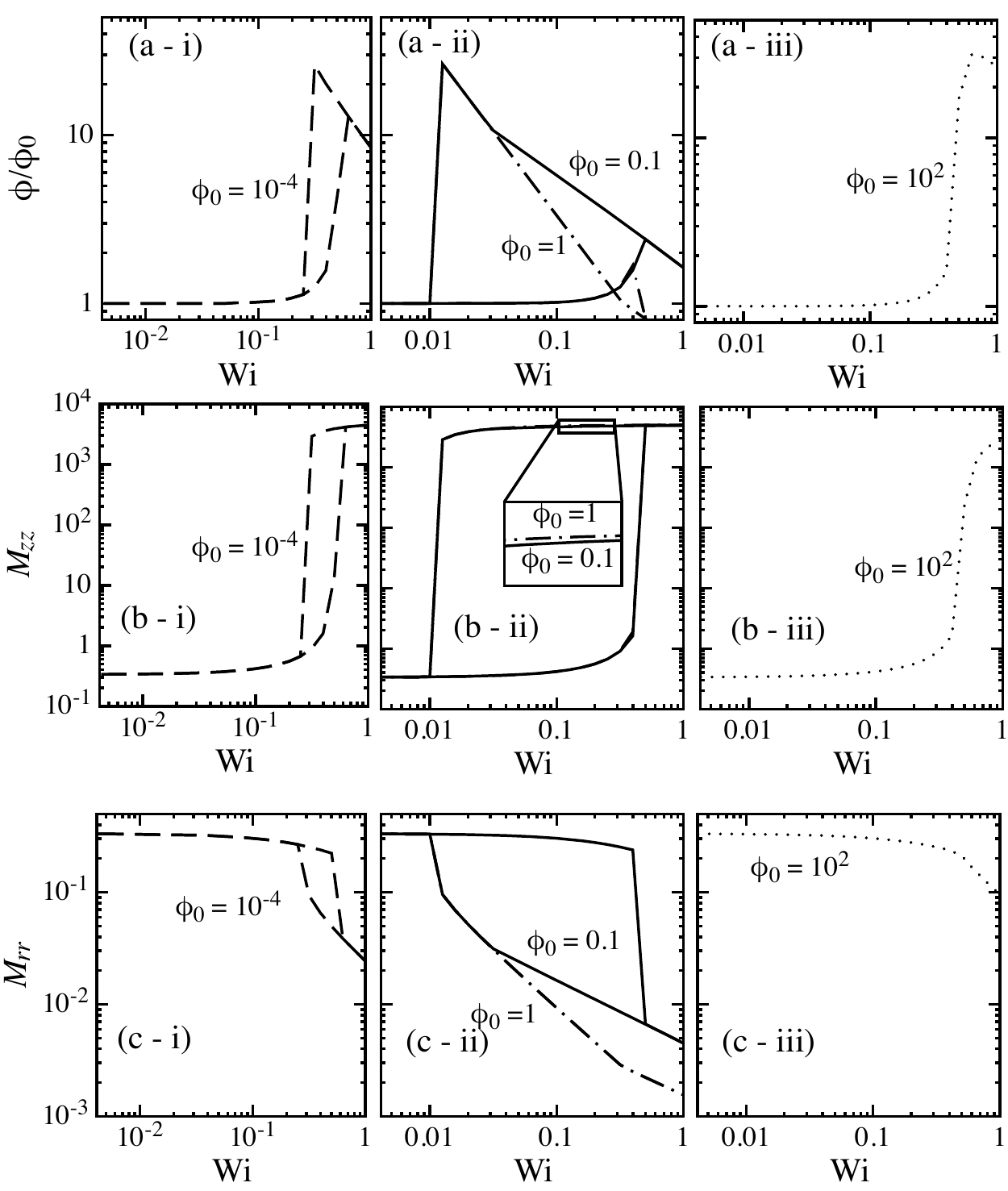}}}
\caption{\label{f:sizes} Coil-stretch hysteresis in (a) pervaded volume fraction ratio $\phi/\phi_0$, (b) $\Mzz$, and (c) $\Mrr$ at (i) low (left column of plots), (ii) intermediate (central column), and (iii) high (right column), concentrations}
\end{figure}

The strong concentration dependence of $\Wics/\Wisc$ for values of $\phio \ll 1$ and the maximum can be understood by considering the influence of the non-equilibrium pervaded volume fraction $\phi$ on the ratio $\zeta_\Upsigma/ \zeta_0$. Figure~\ref{f:sizes} shows steady-state predictions by the CDD model for chain dimensions and $\phi$ for various equilibrium concentrations $\phio$. When there is significant hysteresis, it is clear that the largest non-equilibrium $\phi$ is obtained at the stretched state at the SCT. The values of the largest $\phi/\phio$ ratio in Fig.~\ref{f:sizes} obtained with the CDD model compare well with the approximate estimate,
\begin{gather}
\frac{\phi_\Upsigma}{\phio} =  \frac{\ell_\Upsigma }{d_0} \, \left(\frac{d_\Upsigma}{d_0}\right)^2 = \sqrt{\frac{\NK}{3}} \,,
\label{e:phiscphi0}
\end{gather}
derived in Appendix~\ref{a:stretch}. Thus, pervaded volumes at the SCT determining the friction ratio $\zeta_\Upsigma/\zetac$ and thereby $\Wics/\Wisc$ can be much larger than at equilibrium. 

It is useful at this stage to define critical concentrations for the $\Upsigma$-state. The critical overlap concentration for overlap for equilibrium coils $\phistarc = 1$. Due to the larger molecular volumes in the $\Upsigma$-state, overlap sets in at a lower concentration, which is obtained by setting $\phi_\Upsigma =1$ in the equation above as
\begin{gather}
\phistars =  \left( \frac{\NK}{3} \right)^{-1/2} \,.
\label{e:phistars}
\end{gather}
Further, the critical concentration at which HI screening is complete in equilibrium coils was identified in Eqn.~\eqref{e:phiodag} earlier as $\phidagrc = \sqrt{3 \NK}$. Its counterpart for the $\Upsigma$-state is obtained by substituting in Eqn.~\eqref{e:zetasemidil} the average chain dimensions in $\Upsigma$-state, and equating $\zetar/\zetaZ$ to $\zetaR/\zetaZ = \sqrt{3 \NK}$:  
\begin{equation}
 \phidagrs = \frac{2\sqrt{3}}{\sqrt{\NK}}\,.
\label{e:phiodagger}
\end{equation}
These four critical concentrations --- $\phistars <  \phidagrs <  \phistarc ( = 1) <  \phidagrc$ --- mark key points in the concentration-dependence of $\Wics/\Wics$ in Figs.~\ref{f:wiratio}. 

\begin{figure}[h!]
\centerline{\resizebox{12.7cm}{!}{\includegraphics{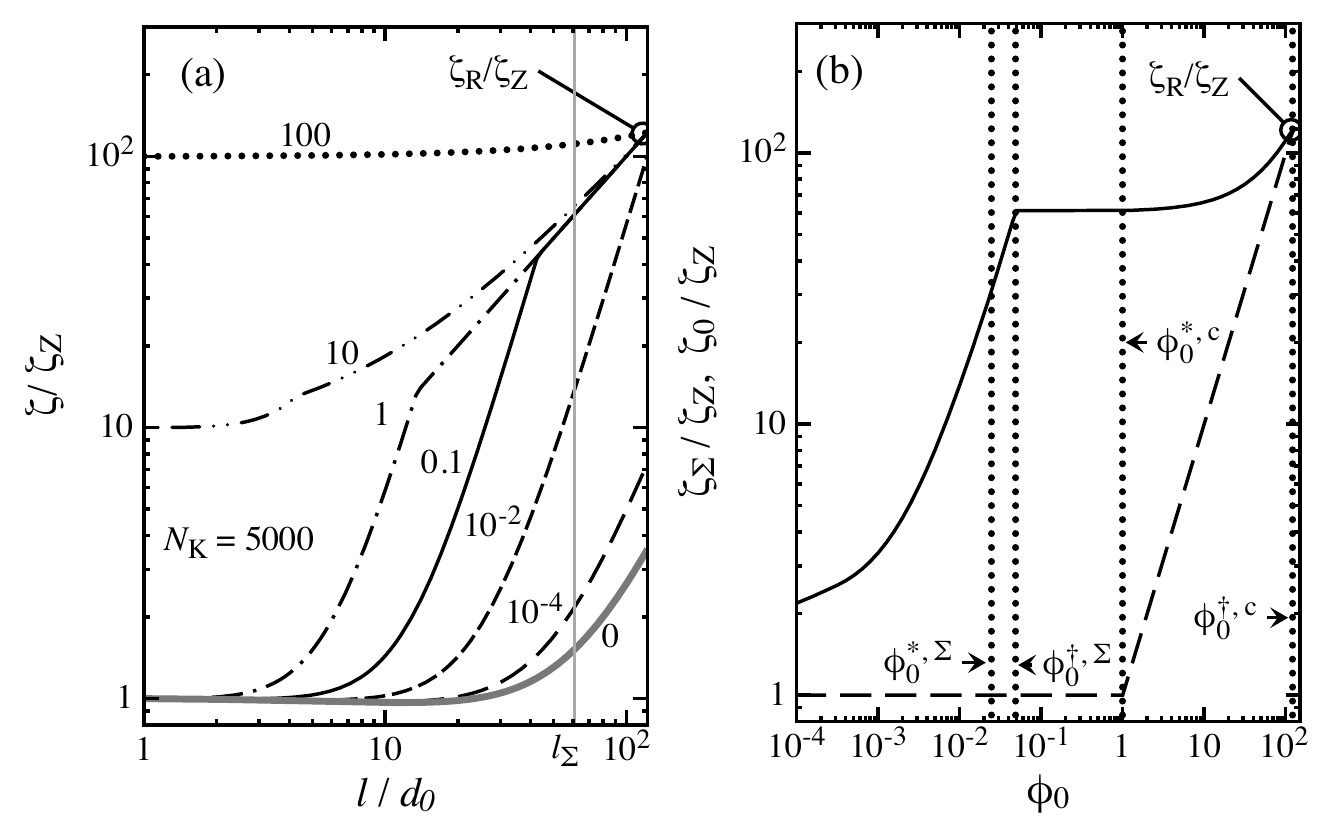}}}
\caption{\label{f:zeta} The friction coefficient ratio $\zeta/\zetaZ$ at a constant transverse chain size of $d_\Upsigma = \sqrt{2/3}\,d_0$: (a) Variation with molecular stretch $\ell$ for various values of $\phio$. The numbers alongside curves in (a) indicate $\phio$, while the vertical line indicates the $\Upsigma$-state with $\ell_\Upsigma = L/2$ (for $\NK = 5000$, $L/(2 d_0) = \sqrt{3 \NK}/2 = 61$). (b) Variation of $\zeta/\zetaZ$ with $\phio$ for stretched chains at the SCT: the dashed curve in (b) shows the variation of $\zetac/\zetaZ$ with $\phio$.   The maximum value for the friction coefficient $\zetaR/\zetaZ$ and $L/d_0$ are both equal to $\sqrt{ 3 \NK}$ ( = 122 for $\NK = 5000$).   }
\end{figure}

To understand changes in $\zeta_\Upsigma$ with $\phio$ at the stretch-coil transition, Fig.~\ref{f:zeta} (a) plots for each $\phio$, the average $\zeta/ \zetaZ$ for a molecule as a function of its stretch with its transverse size held fixed at $d_\Upsigma = \sqrt{2/3}\, d_0$. The grey vertical line indicates the stretch corresponding to $\ell_\Upsigma = L/2$. The change in $\zeta_\Upsigma/ \zetaZ$ with $\phio$ along this line is shown in Fig.~\ref{f:zeta} (b), along with $\zetac/\zetaZ$ modeled by Eqn.~\eqref{e:zetac2}. 

When $\phio < 1$, the coiled-state friction remains unchanged at $\zetac = \zetaZ$. In this concentration range, when $\phio \lesssim \phistars $, the $\Upsigma$-state friction coefficient is under transition from the Batchelor regime. When $\phistars \leq \phio < \phidagrs$, from Eqn.~\eqref{e:zetasemidil},
\begin{gather}
\frac{\zeta_\Upsigma}{ \zetaZ} \approx  \frac{\zeta_\mathrm{r,\,\Upsigma}}{ \zetaZ} = 2\,\phio \, \NK \,; 
\label{e:zetaupsigma-1}
\end{gather}
that is, $\zeta_\Upsigma$ grows linearly with concentration.  Therefore, over the entire range of concentrations $\phio < \phidagrs$, the ratio $\zeta_\Upsigma/ \zetac$ and thus the hysteresis window size $\Wics/\Wisc$ (Eqn.~\eqref{e:wiratio-zetaratio}) increases with concentration, particularly in the range $\phistars \leq \phio < \phidagrs$, where the increase is linear.  The relative width $\phidagrs/\phistars = 6$ of the concentration regime where linear concentration dependence is obtained is quite large and independent of molecular weight. In the range $ \phidagrs \leq \phio < 1$, neither $\zeta_\Upsigma = \zetaR$ nor $\zetac$ change, and therefore $\zeta_\Upsigma/ \zetac$ remains constant leading to a broad plateau in the hysteresis width-versus-concentration plot, spanning a relative width of $1/\phidagrs \sim \sqrt{\NK}$. 

In the nominally semidilute solution ($\phio \geq  1$), the coiled- state friction coefficient $\zetac = \phio \,\zetaZ$ while $\zeta_\Upsigma$ remains effectively constant. (In Fig.~\ref{f:zeta} (b), a relatively small change in $\zeta_\Upsigma$ is observed since it is calculated as a weighted-average of $\zetar$ and $\zetac$, according to Eqn.~\eqref{e:mixrule}.) Therefore, when $1 \leq \phio < \phidagrc$, the ratio  $\zeta_\Upsigma/ \zetac \sim \phio^{-1}$ and the hysteresis window shrinks correspondingly with increasing $\phio$, completely vanishing at $\phio = \phidagrc$ when HI screening is complete in the coiled state as well. Thus, beyond  $\phidagrc$, conformation-dependence of the friction coefficient disappears, and predictions are identical to those obtained with the conventional FENE-P model.

The steady-state results presented above show that self-concentration influences coil--stretch hysteresis over a broad range of concentrations spanning from about $\phistars$ to $\phidagrc$, the relative size $\phidagrc/\phistars$ of this concentration domain being proportional to $\NK$. The height of the maximum in the hysteresis width $\Wics/\Wisc$ over the asymptotic value in the dilute limit given by Batchelor's theory is
\begin{gather}
\frac{\left(\Wics/\Wisc\right)_\textrm{plateau}}{\left(\Wics/\Wisc\right)_\textrm{dil. limit}}\, =  \, \frac{\zetaR}{\zeta_{\Upsigma,\,\textrm{Batchelor}}} \approx \frac{\ln \,\NK}{K} \approx 10 \, \ln \NK  \,.
\label{e:wiratio-plateau}
\end{gather}
While the dependence on molecular weight is only logarithmic, the value of the right-hand side can be large, particularly because of the factor $1/K$. Although the model presented here has ignored other numerical pre-factors, it will be shown in the following section that the estimate above provides an upper bound for the magnitude of the self-concentration effect on coil-stretch hysteresis observed in capillary-thinning experiments.

\subsection{\label{s:capthin} Capillary thinning dynamics}
The results presented in the previous section are for the steady-state in steady uniaxial extensional flows. As mentioned previously, modeling capillary thinning requires coupling the ODEs for conformation tensor components with the stress-balance governing the macroscopic flow at the mid-filament plane in a liquid bridge. The mid-filament flow is further unsteady, and the strain-rate and Weissenberg number are time-dependent. The transient Weissenberg number $\Wip = \edot(t) \, \lamo$. As before, the characteristic conformational scale is $M_0$. The characteristic macroscopic length, time, and stress scales in this case are the initial mid-filament radius $R_0 $ and $\lamv = 6 \etas R_0/ \gamma$ and $\tauc = \gamma/ R_0$. The pre-factor of 6 is retained in the definition of $\lamv$ so that break-up predicted by the stress balance for the Newtonian solvent occurs at a rescaled $t = 1$. 

With these choices, the  dimensionless version of the stress-balance Eqn.~\eqref{e:EHbal} is
\begin{gather}
\frac{1}{R} - \frac{\edot}{2}  +   \NIp  = 0\,.
\label{e:EHbal-d}
\end{gather}
where $\edot$ above is the strain-rate rescaled by $\lamv^{-1}$. Ignoring an $O(1)$ universal pre-factor for solutions of long flexible molecules under theta conditions, the Zimm relaxation time, 
\begin{gather}
\lamZ = \frac{\etas d_0^3}{\kBT} = \etas \frac{\phio}{\nkBT} \,.
\label{e:lamZ}
\end{gather}
Defining the Deborah number based on $\lamZ$: 
\begin{gather}
\De = \frac{\lamZ}{\lamv} = \frac{\lamZ}{6 \etas R_0/ \gamma}\,.
\end{gather}
from Eqn.~\eqref{e:lamZ} above,
\begin{gather}
\frac{\nkBT}{\gamma/ R_0}  = \frac{\phio }{6\,\De} \,.
\label{e:nkBT}
\end{gather}
The polymeric first normal-stress difference from rescaled Eqn.~\eqref{e:Kramers-d} by $\gamma/ R_0$ is thus
\begin{gather}
\NIp = - \frac{\phio}{2\, \De} \, \ \, f \,(\Mzz - \Mrr)\,,
\label{e:Kramers-d-2}
\end{gather}
where $f$, the FENE-P nonlinearity, is given by Eqn.~\eqref{e:fdef-d} as before.  Noting that $\lamo/\lamZ = \zetac/\zetaZ$, Equation~\eqref{e:EHbal-d} can be reorganized using the definitions of $\Wip$ and $\De$ to give
\begin{gather}
\Wip =  \left[ \, \frac{2\, \De}{R} -  \phio \, \,f \,(\Mzz - \Mrr) \,\right]\, \left( \frac{\zetac}{\zetaZ} \right) \,.
\label{e:Wicapthin}
\end{gather}
It is useful to note that the initial $\Wip$ at  $t = 0$ can be determined \textit{a priori}, and is effectively a function  $\De$ and $\phio$:
\begin{gather}
\left.\Wip\right|_{t = 0} =  2\, \De \,\frac{\zetac}{\zetaZ}  = \begin{cases}
2\, \De  \,, \text{ if } \phio \leq 1 \,,\\
2\, \De\, \phio \,, \text{ if } 1 < \phio \leq \phidagrc \,,\\
2\, \sqrt{3\,\NK} \,\De \,, \text{ if } \phio > \phidagrc \,.
\end{cases}\,.\,.
\label{e:Wicapthin-0}
\end{gather}

Treating the Hencky strain as the independent variable rather than time, the system of ODEs to solve for capillary-thinning is then
\begin{align}
\frac{d t}{d \varepsilon} &= \frac{1}{\edot} = \frac{\De}{\Wip} \,, \label{e:dtbyde} \\[3mm]
\frac{d \Mzz}{d \varepsilon} &= 2 \,\Mzz - \frac{1}{\Wip \, (\zeta/\zetac)}\, \left( \,f\, \Mzz - \frac{1}{3} \right) \,,\label{e:mxxode-d-capthn} \\[3mm] 
\frac{d \Mrr}{d \varepsilon} &= - \Mrr - \,\frac{1}{\Wip \, (\zeta/\zetac)}\, \left( \,f\, \Mrr  - \frac{1}{3} \right)\,.
\label{e:myyode-d-capthn} 
\end{align}
The instantaneous dimensionless mid-filament radius is
\begin{gather}
R = \exp ( - \varepsilon/ 2) \,.
\label{e:R-strain}
\end{gather}
Besides $\NK$ and $\phio$, the Deborah number based on the Zimm relaxation time $\De$ appears as an additional dimensionless parameter in the capillary-thinning problem. Given these three parameters, and starting with the initial conditions $t = 0$ and $\Mzz = \Mrr = 1/3$ at $\varepsilon = 0$, at any $\varepsilon \geq 0$ during the forward integration, $R$ is first determined using Eqn.~\eqref{e:R-strain}.  Since $\Mzz$ and $\Mrr$ is also known at every instant in the integration, the FENE nonlinearity $f$, the instantaneous pervaded volume fraction $\phi$ and the friction ratios $\zeta/\zetac$ and $\zetac/\zetaZ$ are determined as previously described. The transient Weissenberg number $\Wip$ is then  calculated through Eqn.~\eqref{e:Wicapthin} to be substituted on the right-hand side of the ODEs above. These are integrated forward until a large terminal value of $\varepsilon$ (typically, 20). The transient polymer contribution to the extensional viscosity is rescaled by $\nkBT\, \lamo$ as before, and hence:
\begin{gather}
\eetap = - \frac{\NIp}{\Wip} \,.
\end{gather}
In order to compare this with the solvent contribution ($ = 3 \etas$), the latter is also expressed in terms of the same viscosity scale. From Eqn.~\eqref{e:nkBT} above and Eqn.~\eqref{e:zetac2} earlier for the ratio $\zetac/\zetaZ = \lamo/\lamZ$,
\begin{gather}
\eetas = \frac{ 3 \,\etas}{\nkBT \,\lamo} = \frac{3}{\phio \, (\zetac/\zetaZ)}  = 3\,\begin{cases}
\phio^{-1}  \,, \text{ if } \phio \leq 1 \,,\\
\phio^{-2} \,, \text{ if } 1 < \phio \leq \phidagrc \,,\\
\phio^{-1}/\sqrt{3\,\NK} \,, \text{ if } \phio > \phidagrc \,.
\label{e:eetas}
\end{cases}\,.
\end{gather}

 \begin{figure}[h!]
\centerline{{\includegraphics{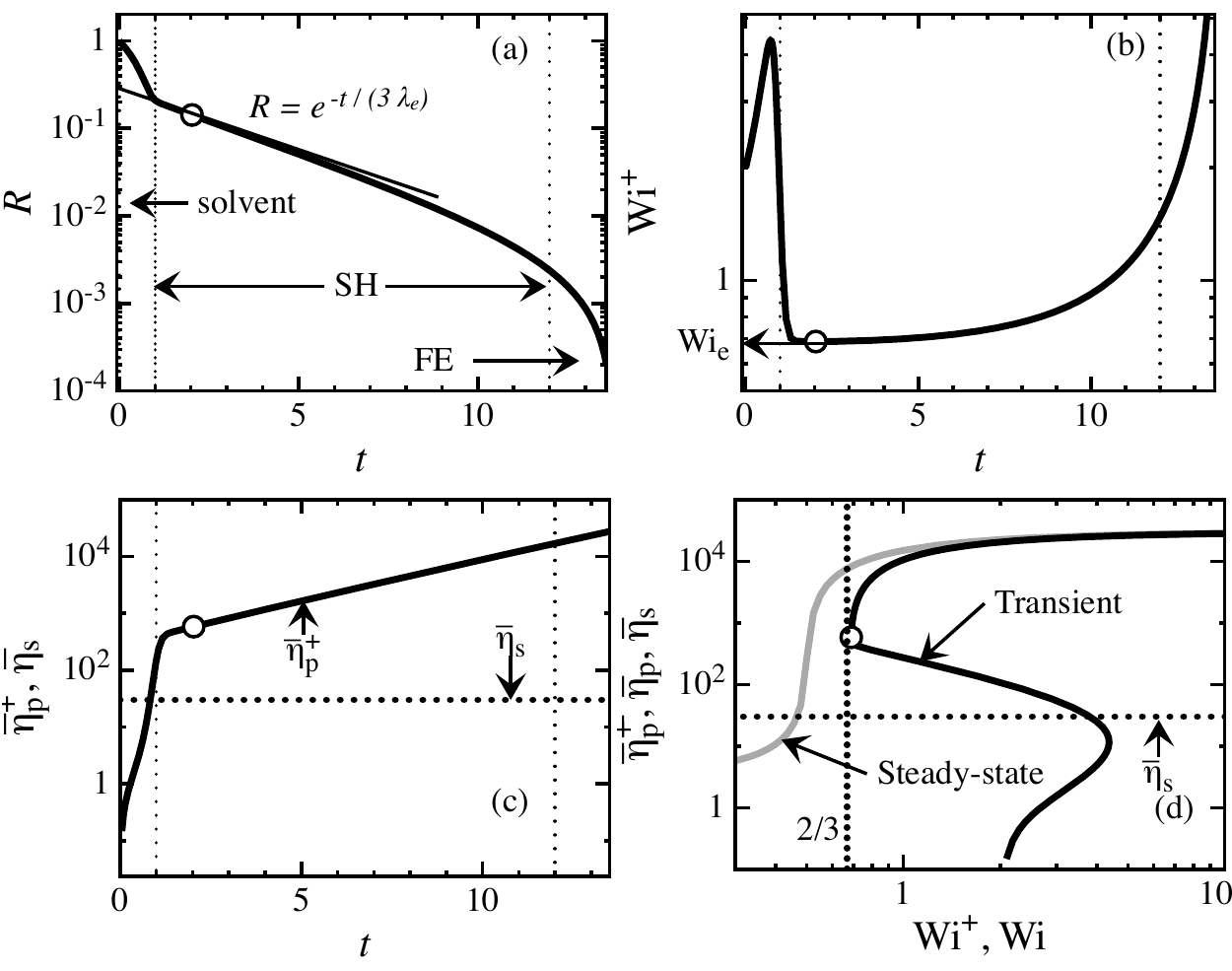}}}
\caption{\label{f:lameextract} Features of  capillary-thinning predictions for dilute polymer solutions with the conventional FENE-P model: (a) mid-filament radius $R$; (b) Weissenberg number, $\Wip$; (c) transient (rescaled) extensional viscosities; (d) comparison of transient $\eetapp$-vs-$\Wip$ with steady-state $\eetap$-vs-$\Wi$. Open symbols represent the point for determining $\Wie$ and $\lame$.  Vertical dotted lines in (a), (b) and (c) demarcate the solvent-, SH- and FE-dominated regimes, In (d), the dotted vertical line is the Entov-Hinch prediction of $\Wie = 2/3$. In (c) and (d), the dotted horizontal line is the rescaled solvent contribution to the extensional viscosity, $\eetas = 3 \,\etas$ (Eqn.~\eqref{e:eetas}).}
\end{figure}

Figure~\ref{f:lameextract} summarizes some well-known features of predictions of the conventional FENE-P model for a typical dilute polymer solution. Initially, polymeric stresses are much smaller than the contribution from the highly viscous solvent and $R$ is predicted to decrease linearly in time. A linear decrease in $R$ implies an exponential increase in $\Wip$. For $\De \gtrsim 1$ in a dilute solution, the initial Weissenberg number (Eqn.~\eqref{e:Wicapthin-0}) is further above $\Wics$. This causes a rapid, nearly exponential, growth in the polymer contribution to the extensional viscosity due to SH  in the initial solvent-dominated phase, and $\eetap$ soon outstrips the solvent contribution. The onset of the polymer-dominated phase is marked by a sudden fall in $\Wip$. In predictions for dilute solutions, the sharp decrease in $\Wip$ presents as a kink in the $R$-vs-$t$ curve (Fig.~\ref{f:lameextract} (a)). As long as the growth in polymeric stress due to SH continues and almost completely balances the growth in capillary stress (as $R^{-1}$), $\Wip$ remains small and may change slowly with time.

Figure~\ref{f:lameextract} (d) combines the data in (b) and (c)  to compare the transient $\eetapp$-vs-$\Wip$ (bold curve) during capillary-thinning against the \textit{steady-state} predictions for $\eetap$ (grey curve) for any value of $\Wi$. The steady-state curve can be regarded as the infinite strain envelope limiting the capillary thinning curve, and the vertical difference between the two curves at any particular value of $\Wi$ indicates the gap that can potentially be filled by SH. As SH progresses, FE becomes important, and subsequently, it is not possible to sustain an increase in polymer viscosity by  an increase in strain at fixed Weissenberg number; instead, $\eetapp$ grows by tracking closely along the steady-state $\eetap$-vs-$\Wi$ curve (Fig.~\ref{f:lameextract} (d)). In other words, the polymer solution behaves rather as  a generalized Newtonian fluid with a viscosity that depends solely on strain-rate  than one that strain-hardens. Thereafter, $\Wip$ must increase for flow-induced  stresses to balance the capillary stress. Eventually, the polymer solution viscosity approaches a constant value,  leading once more to linear  decay in $R$ and exponential growth in $\Wip$ till the filament breaks up.

As Fig.~\ref{f:lameextract} (b)  shows, even with the conventional FENE-P model with a fairly large value of $\NK$, $\Wip$ is not exactly constant during the strain-hardening phase, and therefore, the decay in $R$ is only approximately exponential. The procedure adopted here to define an effective relaxation time $\lame$ during this phase is equivalent to drawing a straight line with the least slope tangential to the strain-hardening portion of the $R$-vs-$t$ curve on a semi-log plot. The minimum value of $\Wip$ in the strain-hardening phase is first identified as $\Wie$. Recalling that the Weissenberg number in this study is based on $\lamo$, and since the tangential straight line is described by $R = e^{-t/(3 \lame)}$, the value of $\lame$ for a model prediction is determined as
\begin{gather}
\frac{\lame}{\lamo} = \frac{2/3}{\Wie} \,; \quad \frac{\lame}{\lamZ} = \left(\frac{\lamo}{\lamZ}\right)\, \frac{2/3}{\Wie} \, \,.
\label{e:lame-wie}
\end{gather}

The results below are all for $\NK = 5000$ and $\De = 1$, which are representative of the polymer solutions used by \citet{clasenetal}. Qualitative features of the effect of polymer concentration are discussed first by comparing predictions of the CDD and FENE-P models for five different values of $\phio$ --- $10^{-4}$, 0.1, 1, 10 and 100. For reference, the critical concentrations for $\NK = 5000$ are: $\phistars = 0.025$, $\phidagrs = 0.049$, $\phistarc = 1$, and $\phidagrc = 122$.  

\begin{figure}[h!]
\centerline{{\includegraphics{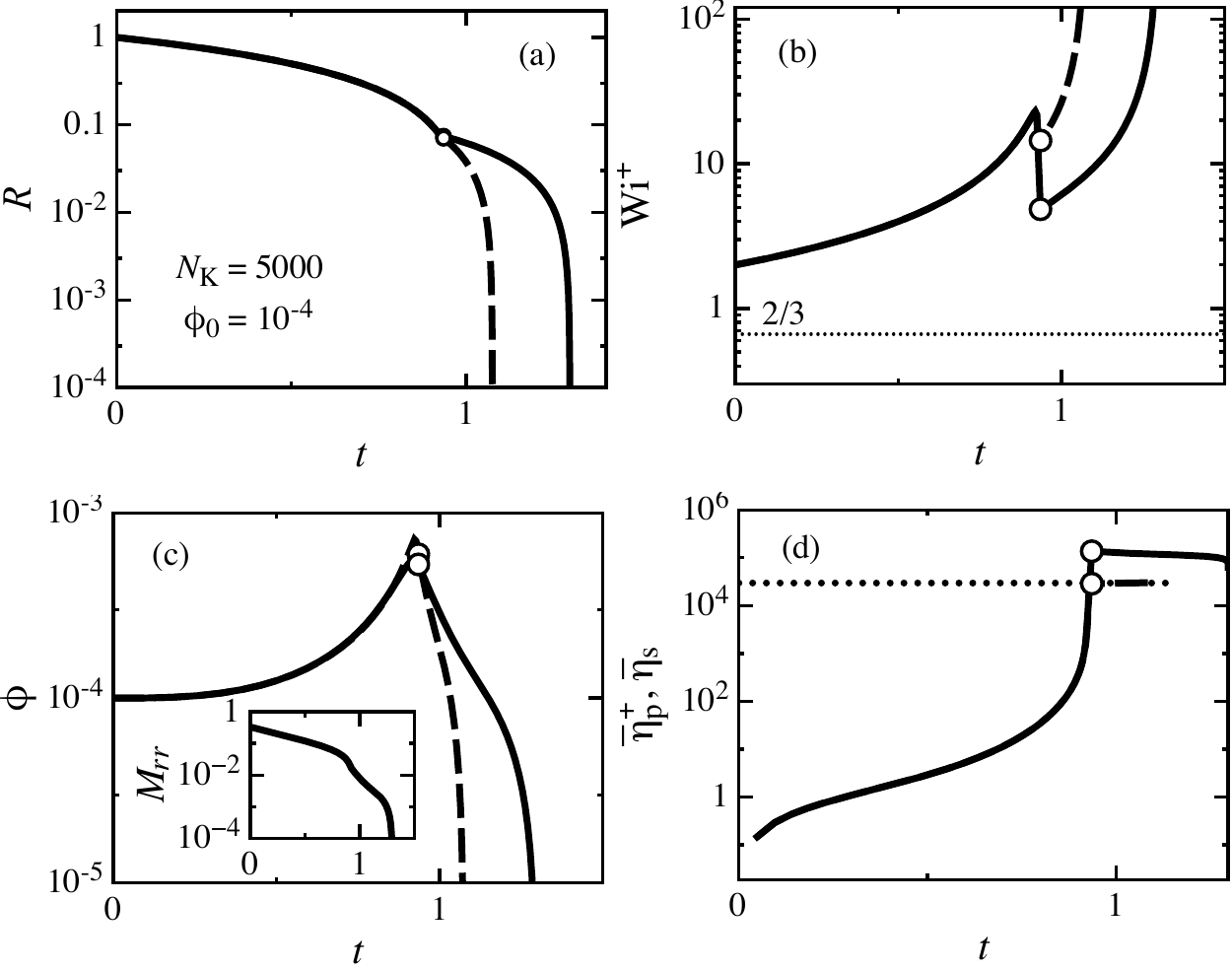}}}
\caption{\label{f:Rvst-1e-4} Capillary-thinning dynamics at $\phi_0 = 10^{-4}$: (a) log-linear plot of mid-filament radius $R$ versus $t$; (b) transient Weissenberg number, $\Wip$; (c) instantaneous volume fraction, $\phi$; (d) transient polymer extensional viscosity, $\eetapp$. The bold continuous and dashed lines are predictions of the CDD and FENE-P models, respectively.The dotted horizontal lines in (b) and (d) represent  the Entov-Hinch prediction of $\Wi = 2/3$ for the elastic regime, and the dimensionless solvent viscosity (scaled by $\nkBT \lambda_0$).  The inset in (c) shows the decrease of the transverse conformation tensor component $\Mrr$ predicted by the CDD model.}
\end{figure}

\begin{figure}[h!]
\centerline{{\includegraphics{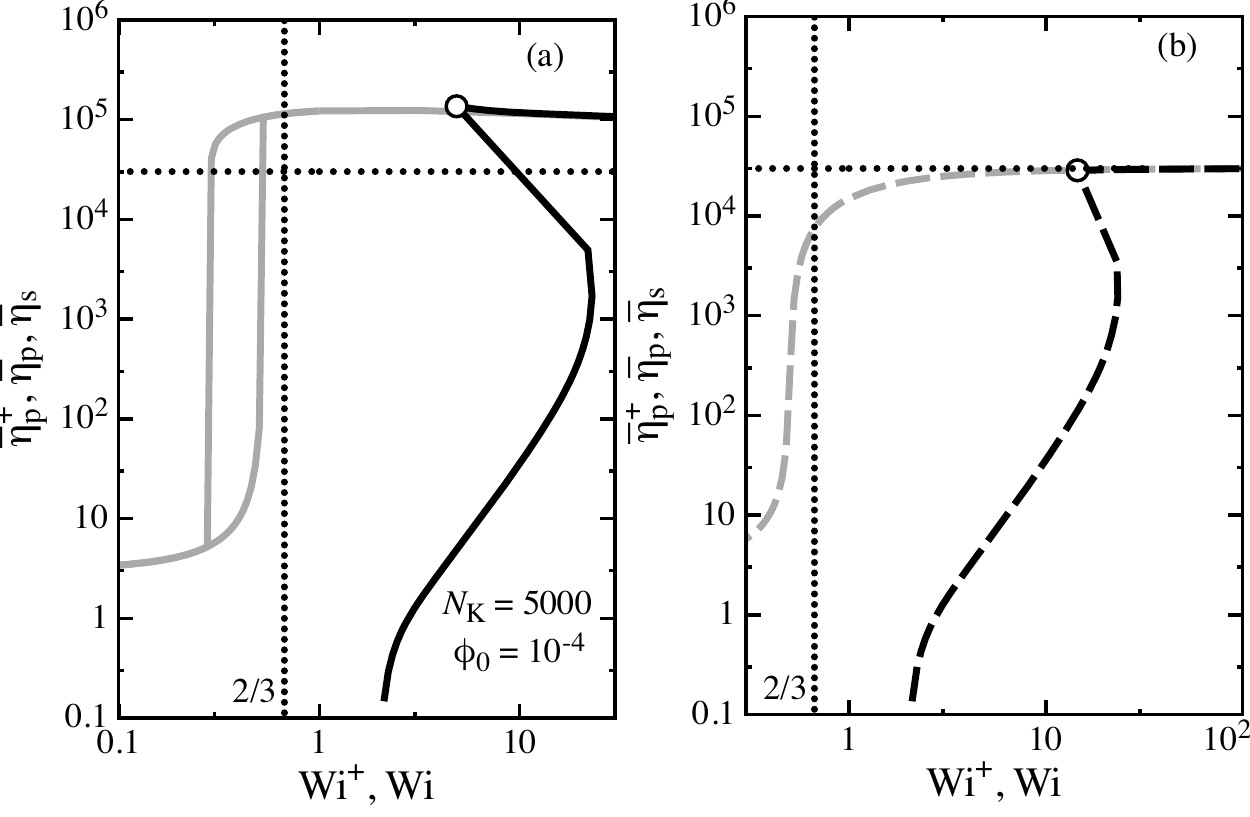}}}
\caption{\label{f:etapvswi-1e-4} Comparison of transient $\eetapp$-vs-$\Wip$ variation during capillary-thinning (bold curves) with steady-state $\eetap$-versus-$\Wi$ (grey curves) predictions of  (a) CDD, and (b) FENE-P models, for $\phio = 10^{-4}$: the dotted vertical and horizontal lines are as in Fig.~\ref{f:lameextract} (d). }
\end{figure}

Figure~\ref{f:Rvst-1e-4} shows the transient changes in $R$, $\Wip$, $\phi$ and $\eetapp$ with  time $t$ for a solution of very low concentration $\phio = 10^{-4} \ll \phistars$. During the solvent-dominated phase of thinning when molecules are not significantly stretched, predictions of the FENE-P and CDD models are indistinguishable from each other (at the resolution in the plots). By the time the SH phase sets in and $\Wip$ falls, stretching is large enough for predictions of the two models to deviate from one another.  At the low concentration in Fig.~\ref{f:Rvst-1e-4}, the crossover to the SH phase occurs quite late since the polymer-induced stress per molecule must be large enough that its product with the low concentration is larger than the solvent contribution.  As a result, both models predict an almost immediate transition to the FE dominated regime, and no distinct SH regime with a significant plateau in $\Wip$ is observed in Fig.~\ref{f:Rvst-1e-4} (b). The brief minimum in $\Wip$ is however clear, leading to kinks in the $R$-vs-$t$ curves for either model. The minimum values of $\Wip$ are considerably larger than $2/3$, leading to values of $\lame$ \textit{much smaller than} $\lamo = \lamZ$ (Eqn.~\eqref{e:lame-wie}). Figure~\ref{f:etapvswi-1e-4} shows the transient $\eetapp$-vs-$\Wip$ variation relative to the steady-state results of the two models.

The instantaneous pervaded volume fraction $\phi$ with either model is interestingly predicted to first increase substantially in the solvent-dominated phase (Fig.~\ref{f:Rvst-1e-4} (c)), even though $\Wip$ is quite large.  As the inset in at the Fig.~\ref{f:Rvst-1e-4} (c) shows, although $\Mrr$ and hence the transverse size decrease in this phase, the growth in $\Mzz$ (not shown) is more than enough to compensate and lead to an increase in $\phi$. The levelling-off of $\eetapp$ in Fig.~\ref{f:Rvst-1e-4} (d) indicates that,  in the FE dominated regime, end-to-end stretch stabilizes; the transverse size continues to shrink with increasing $\Wip$ (inset in Fig.~\ref{f:Rvst-1e-4} (c)). As a result, chain volume fraction in the terminal regime rapidly decreases to values much lower than the equilibrium value initially. 

Another notable feature of the FENE-P prediction in  Fig.~\ref{f:Rvst-1e-4} (d) is that the FE-limited value of $\eetapp$ is about the same as $\etas$. This implies that for much lower concentrations, the solvent contribution will be dominant even after the polymer contribution attains the FE limited plateau; in other words, the influence of the polymer will not be significant and the radial decay will be effectively identical to that of the pure solvent until break-up. As \citet{clasenetal} pointed out, it is possible to calculate from the asymptotic steady-state $\eetap$ at full molecular stretch the value of the minimum polymer concentration required to observe a significant viscoelastic response during capillary-thinning. 

\begin{figure}[h!]
\centerline{{\includegraphics{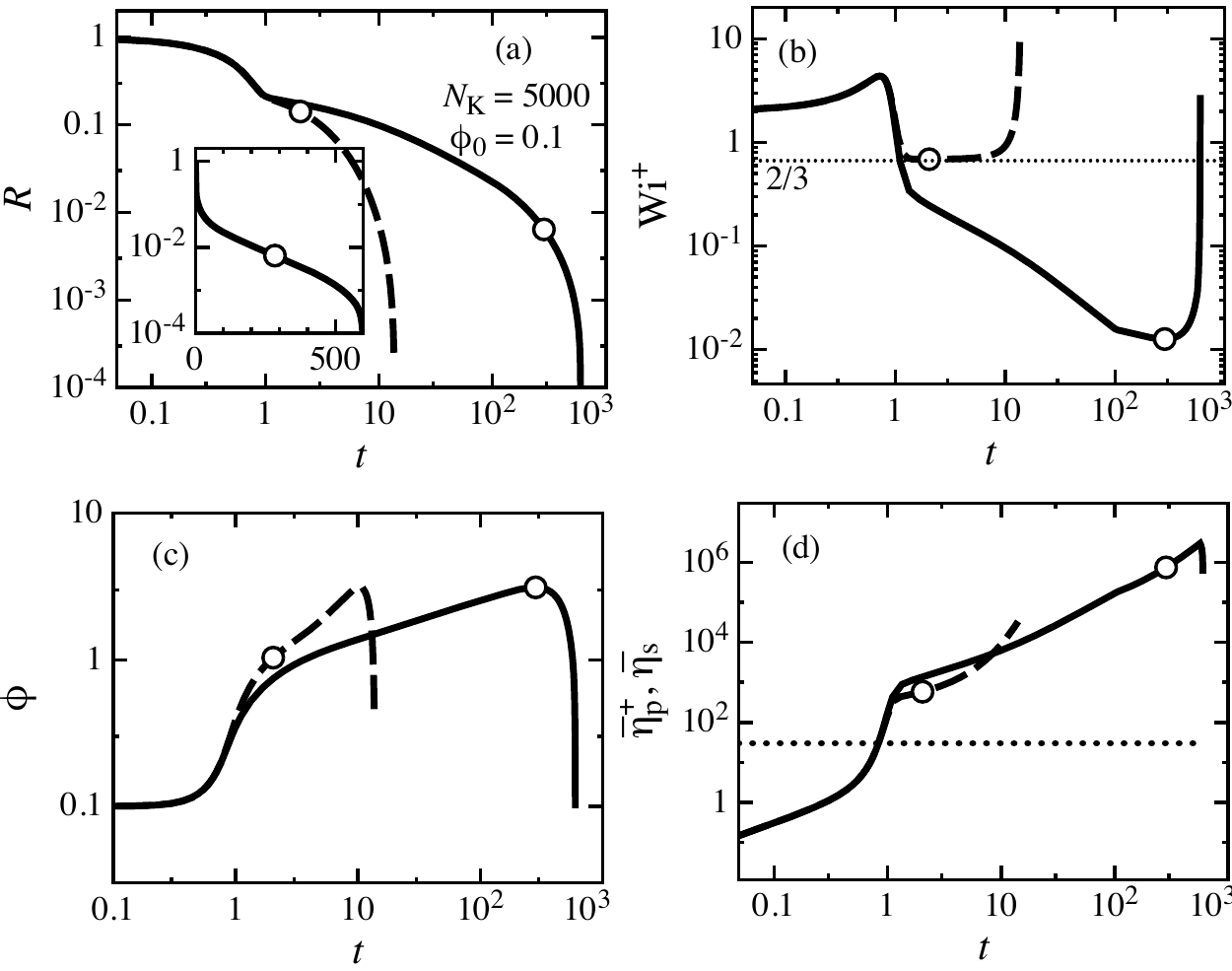}}}
\caption{\label{f:Rvst-1e-1} Capillary-thinning dynamics at $\phio = 0.1$: (a) log-log plot of mid-filament radius $R$ versus $t$ (inset shows log-linear plot); (b) transient Weissenberg number, $\Wip$; (c) instantaneous volume fraction, $\phi$; (d) transient polymer extensional viscosity, $\eetapp$; symbols are as in Fig.~\ref{f:Rvst-1e-4}.}
\end{figure}

Several qualitative differences emerge between the predictions of the two models progressively with increasing concentration. These are exemplified by the transient behaviour at $\phio = 0.1$ in Fig.~\ref{f:Rvst-1e-1}. The overall break-up time predicted by the CDD model becomes larger by almost two orders of magnitude. The SH phase predicted by both models is more pronounced than that observed previously at $\phio = 10^{-4}$ (Fig.~\ref{f:Rvst-1e-1} (a) and (b)). A clear plateau of $\Wip$ emerges with $\Wie = 2/3$ in the prediction of the FENE-P model. Interestingly however, no such plateau occurs (on a logarithmic axis) with the CDD model; instead, after its sharp fall following the onset of the polymer-stress dominated SH regime, $\Wip$ continues to fall \textit{well below the value of 2/3}, attaining a very low minimum $\Wie$ value of about $10^{-2}$. Although this is corresponds to a very low value of the strain rate, the polymer viscosity $\eetapp$ (Fig.~\ref{f:Rvst-1e-1} (d)) attained  at the point of minimum $\Wie$ is considerably greater than the value attained by the FENE-P model for the same strain or $R$ value of about $10^{-2}$. To achieve the balance against the capillary stress corresponding to this value of $R$, the FENE-P model is forced to compensate for the lower value of $\eetap$ by attaining a higher $\Wip$. The instantaneous volume fraction (Fig.~\ref{f:Rvst-1e-1} (c))in the SH regime is well beyond overlap. This self-concentration in the CDD model permits a very large extensional polymer viscosity --- and  therefore, a highly  stretched average conformation --- to be sustained at very low $\Wip$.  

\begin{figure}[h!]
\centerline{{\includegraphics{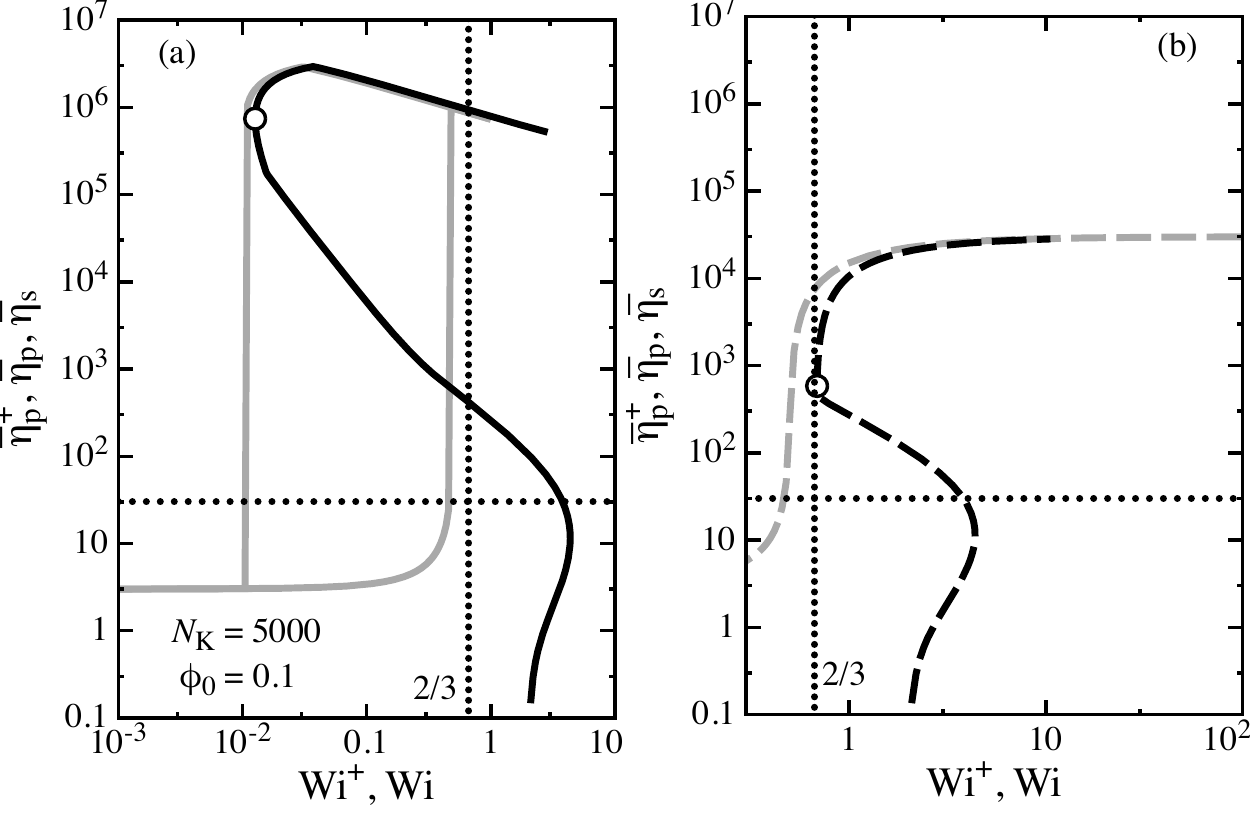}}}
\caption{\label{f:etapvswi-1e-1} Comparison of transient $\eetapp$-versus-$\Wip$ variation during capillary-thinning with steady-state $\eetap$-versus-$\Wi$ predictions of  (a) CDD, and (b) FENE-P models, for $\phio = 0.1$: symbols are as in Fig.~\ref{f:etapvswi-1e-4}.   }
\end{figure}

Figure~\ref{f:etapvswi-1e-1} compares the transient $\eetapp$-vs-$\Wip$ curves (bold curves in (a) and (b)) for the CDD and FENE-P models against their corresponding steady-state predictions (grey curves). The concentration of $\phio = 0.1$ lies between the critical concentrations of $\phidagrs = 0.049$ and $\phistars = 1$. Thus, the size of the steady-state hysteresis window predicted by the CDD model at this concentration is at its maximum (continuous grey curve in Fig.~\ref{f:etapvswi-1e-1} (a)). \textit{It is seen that the value of $\Wie$ attained with the CDD model is essentially the same as the stretch--coil transition $\Wisc$}; in sharp contrast, $\Wie$ with the FENE-P model is 2/3. The corresponding predictions of $\lame$ with the CDD model is therefore \textit{considerably greater} than the value of $\lamo$ at the same $c/\cstar = \phio$ (Eqn.~\eqref{e:lame-wie}), whereas with the FENE-P model, $\lame = \lamo$ at this concentration. 

\begin{figure}[h!]
\centerline{{\includegraphics{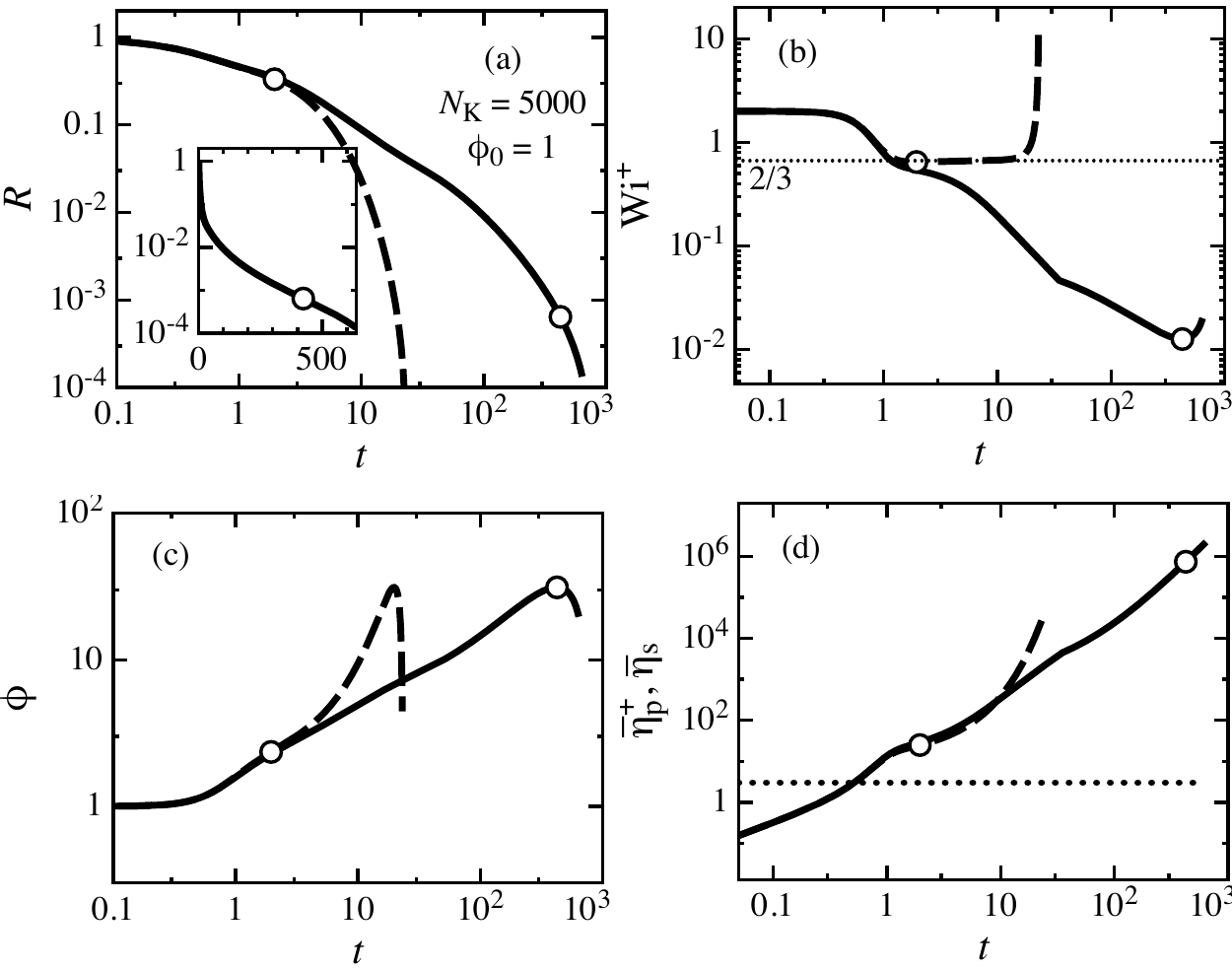}}}
\caption{\label{f:Rvst-1} Capillary-thinning dynamics at $\phio = 1$: (a) log-log plot of mid-filament radius $R$ versus $t$ (inset shows log-linear plot); (b) transient Weissenberg number, $\Wip$; (c) instantaneous volume fraction, $\phi$; (d) transient polymer extensional viscosity, $\eetapp$; symbols are as in Fig.~\ref{f:Rvst-1e-4}.}
\end{figure}

When concentration is increased further to $\phio = 1$, the solution is at the edge of the semidilute domain (Fig.~\ref{f:Rvst-1}). The initial polymer contribution to the flow-induced stresses is no longer negligible in comparison with the solvent contribution. Therefore, there is no clearly distinguishable initial solvent-dominated phase. Initially, $\Wip$ is nearly constant as the polymer contribution also plays an equal role in balancing the capillary stress. Since $\Wip > \Wics$, rapid stretching occurs and as $\eetapp$ increases $\Wip$ falls. With the FENE-P, $\Wip$ does not fall below $2/3$, but instead plateaus at that value. With the CDD model, on the other hand, $\Wip$  continues to falls in the SH regime, again reaching a very low $\Wie \ll 2/3$ (Fig.~\ref{f:Rvst-1} (b)), the large $\eetapp$ being the result of self-concentration (Fig.~\ref{f:Rvst-1} (c)).

\begin{figure}[h!]
\centerline{{\includegraphics{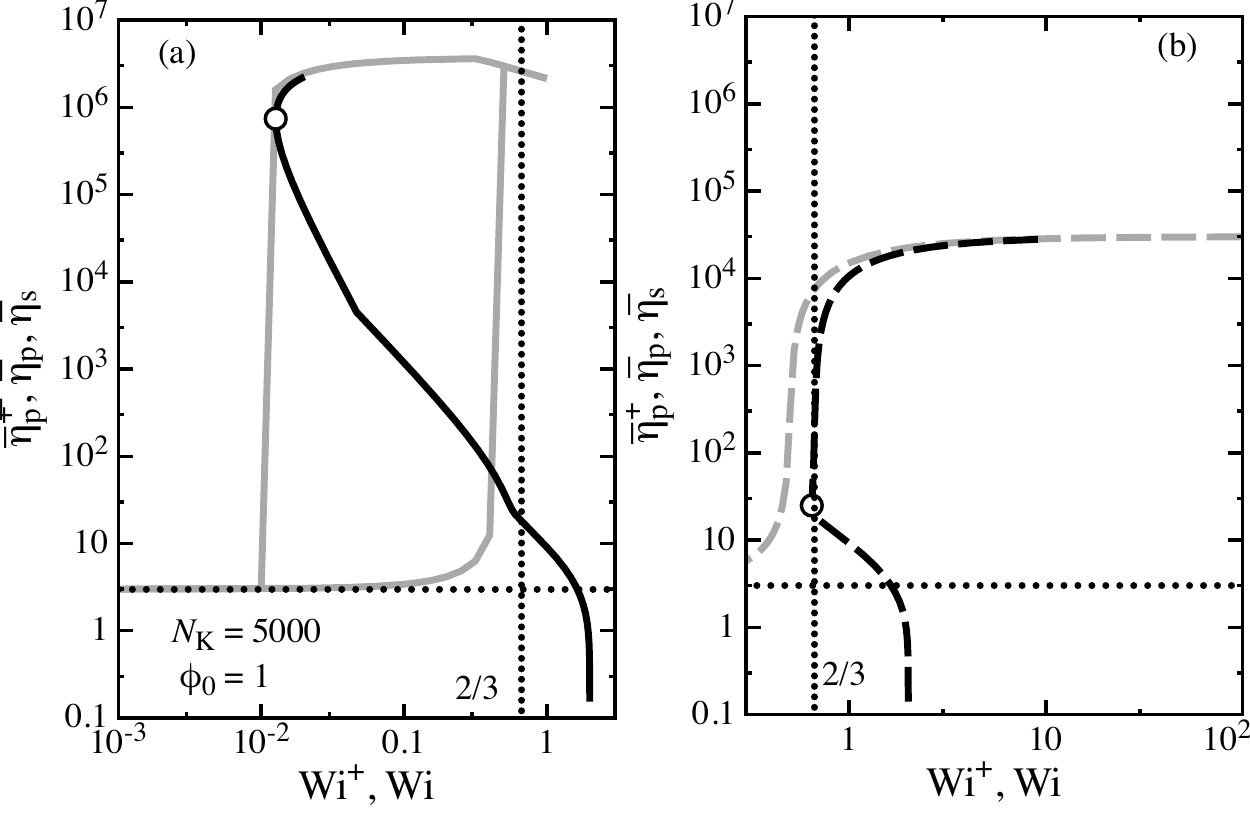}}}
\caption{\label{f:etapvswi-1} Comparison of transient $\eetapp$-versus-$\Wip$ variation during capillary-thinning with steady-state $\eetap$-versus-$\Wi$ predictions of  (a) CDD, and (b) FENE-P models, for $\phio = 1$: symbols are as in Fig.~\ref{f:etapvswi-1e-4}.   }
\end{figure}

Many features of the $\eetapp$-vs-$\Wie$ plots in Fig.~\ref{f:etapvswi-1} for $\phio = 1$ are similar to those earlier for $\phio = 0.1$ in Fig.~\ref{f:etapvswi-1e-1} , particularly the fact that, as before $\Wie$ in the FENE-P model is bound by the Entov-Hinch limit fo $2/3$ whereas, in the CDD model, $\Wie = \Wisc$. The similarities between predictions of the CDD model for $\phio = 0.1$ and $1$ are due to the similarities between the steady-state hysteresis loops in $\eetap$ in the two cases. These two concentrations lie in the domain between $\phidagrs$ and $\phistarc$ in which the hysteresis window size remains unchanged (Fig.~\ref{f:wiratio}). Thus, as at $\phi = 0.1$, $\lame \gg \lamo = \lamZ$ with the CDD model, while $\lame = \lamo$ with the FENE-P model.


\begin{figure}[h!]
\centerline{{\includegraphics{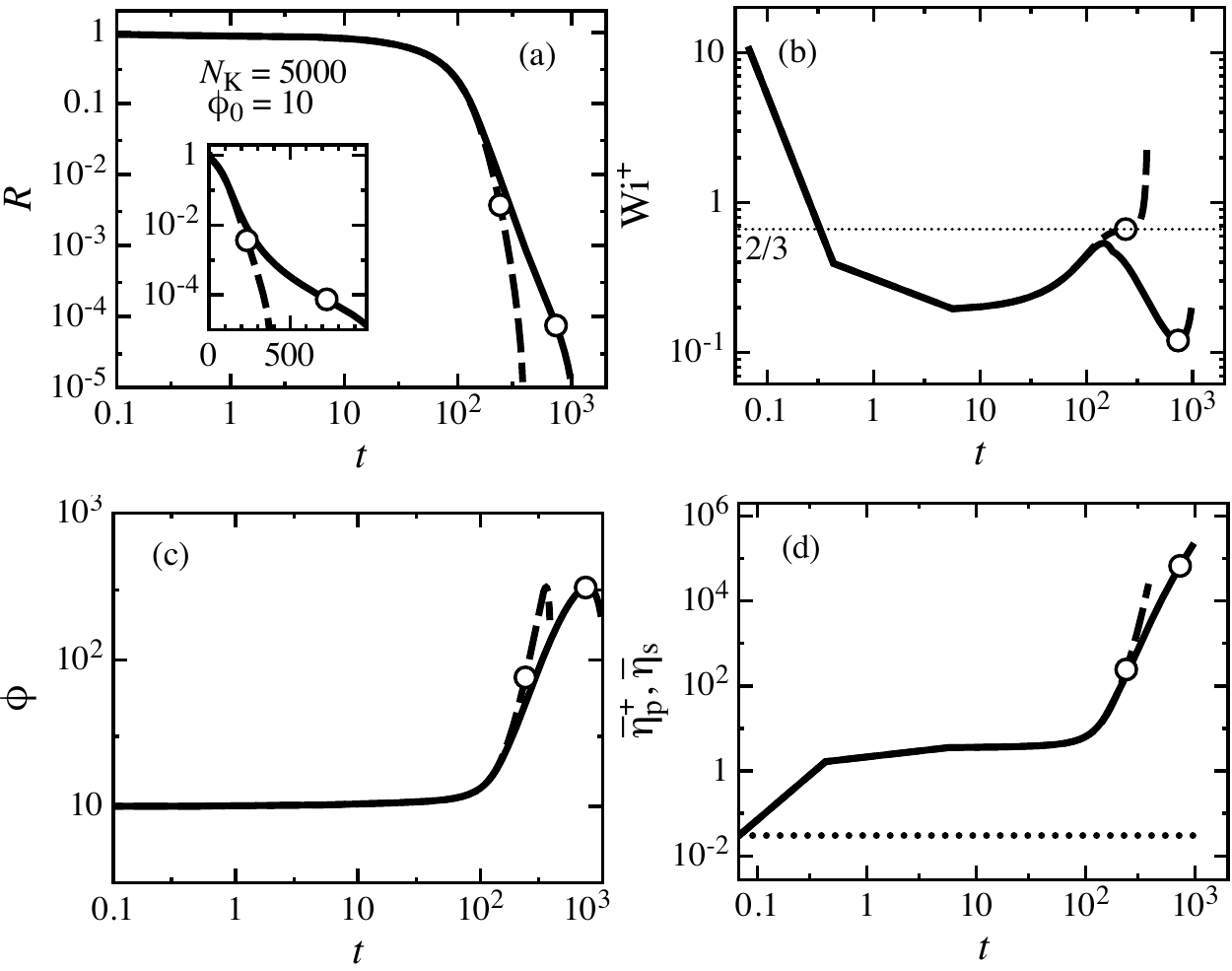}}}
\caption{\label{f:Rvst-10} Capillary-thinning dynamics at $\phi_0 = 10$: (a) log-log plot of mid-filament radius $R$ versus $t$ (inset shows log-linear plot); (b) transient Weissenberg number, $\Wip$; (c) instantaneous volume fraction, $\phi$; (d) transient polymer extensional viscosity, $\eetapp$; symbols are as in Fig.~\ref{f:Rvst-1e-4}.}
\end{figure}

In semidilute solutions with $\phio \gg 1$, the solvent contribution plays an insignificant role, even initially. As illustrated by the behaviour at $\phio = 10$ in Fig.~\ref{f:Rvst-10}, the initial filament evolution  is different from the rapid initial linear thinning predicted in the case of dilute solutions, with $\Wip$ falling initially and increasing slowly, but remaining at low values much smaller than $\Wics$. Since the polymer concentration is large, the initially low capillary stress at large $R \lesssim 1$ can be completely balanced by $\eetapp$ comparable to the rescaled zero-strain-rate extensional viscosity $\eetapo \approx 3$. Since the dominant $\eetapp$ is nearly constant, the radial decay is nearly linear initially (Fig.~\ref{f:Rvst-10} (a) inset).

\begin{figure}[h!]
\centerline{{\includegraphics{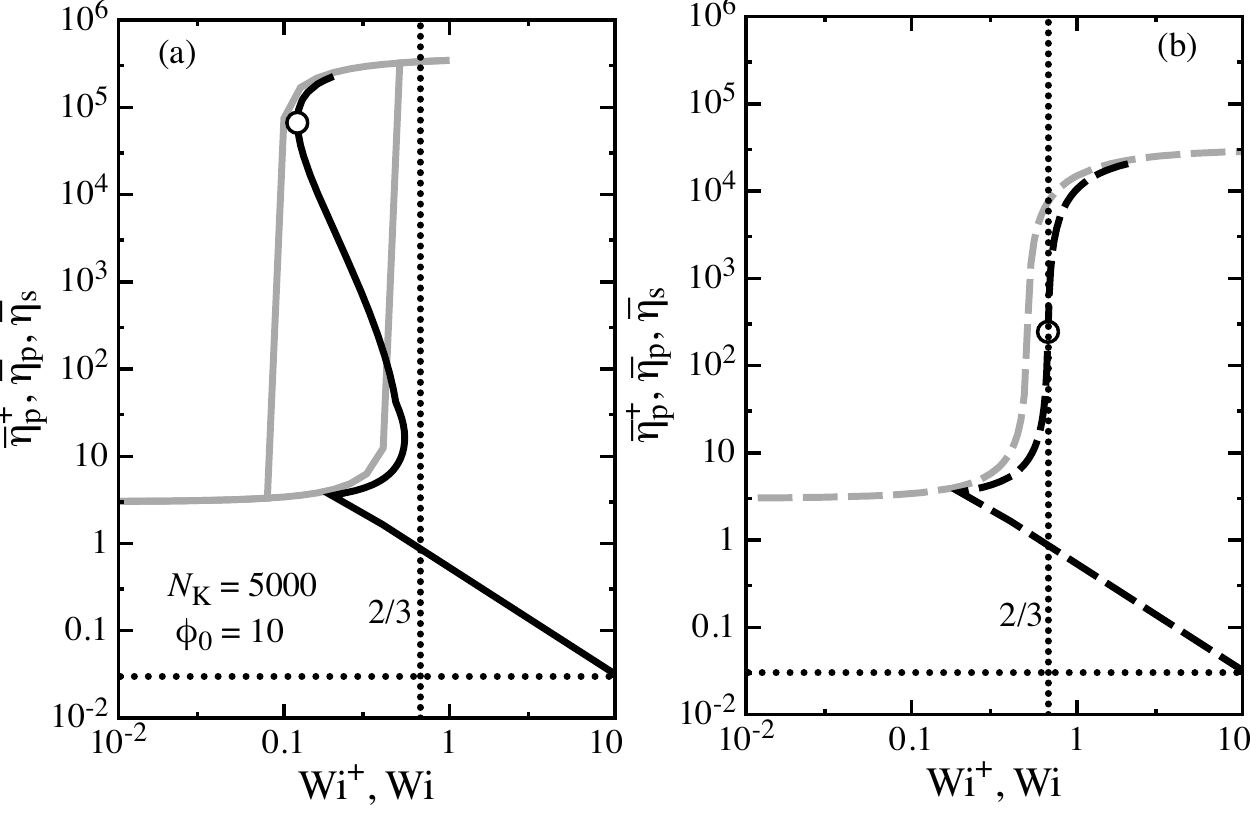}}}
\caption{\label{f:etapvswi-10} Comparison of transient $\etap^+$-versus-$\Wi^+$ variation during capillary-thinning with steady-state $\eetap$-versus-$\Wi$ predictions of  (a) CDD, and (b) FENE-P models, for $\phi_0 = 10$: symbols are as in Fig.~\ref{f:etapvswi-1e-4}.   }
\end{figure}

In this initial phase, the transient $\eetapp$ predicted by both the models tracks along the steady-state $\eetap$ for the \textit{coiled} state until $\Wip$ increases beyond $\Wics = 1/2$  (Fig.~\ref{f:etapvswi-10}). Thereafter, the SH regime sets in, and  $\Wip$ and $\eetapp$ grow relatively rapidly in the FENE-P model. In this case, the Entov-Hinch value is approached \textit{from below} (Figs.~\ref{f:Rvst-10} (b) and \ref{f:etapvswi-10} (b)) in the FENE-P prediction, and instead of a broad minimum as observed in the previous cases, an inflection point occurs at $\Wip = 2/3$  (Fig.~\ref{f:Rvst-10} (b)). Since the effective time-constant $\lame$ is defined in the SH regime, this inflection point is chosen as $\Wie$. Thus, as before, $\lame  = \lamo$ with the FENE-P model, but since $\phio > 1$, $\lamo = \phio \, \lamZ$ in this case. With the CDD model, the steady-state hysteresis loop causes $\Wip$ to fall again after it ``rounds the bend" beyond $\Wics = 1/2$ while $\eetapp$ continues to increase in the SH phase (Fig.~\ref{f:etapvswi-10} (a)). The strain-rate drops until a minimum at $\Wie = \Wisc$ again, before the system enters the FE limited regime, and $\Wip$ grows. Since in this case $\phistarc < \phio < \phidagrc$, the hysteresis window is significantly smaller than at $\phio = 1$. The effective relaxation time $\lame
> \lamo$ again, but the ratio $\lame/\lamo$ is smaller than its value at either $\phio = 0.1$ or $1$. Due to the smaller hysteresis window, despite the clear qualitative differences between the two models, the $R$-vs-$t$ curves in Fig.~\ref{f:Rvst-10} (a) are quite similar.

\begin{figure}[h!]
\centerline{{\includegraphics{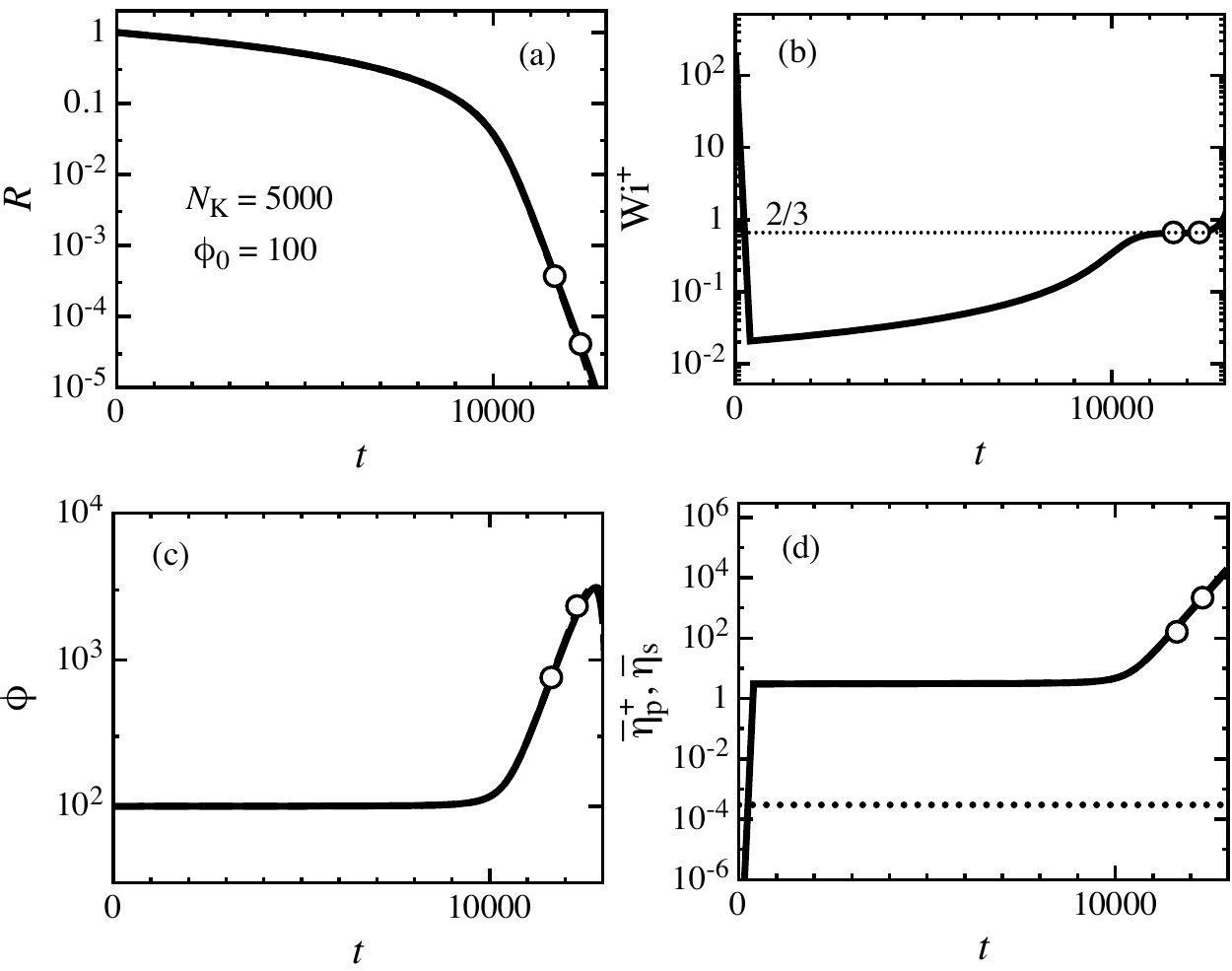}}}
\caption{\label{f:Rvst-100} Capillary-thinning dynamics at $\phi_0 = 100$: (a) log-log plot of mid-filament radius $R$ versus $t$; (b) transient Weissenberg number, $\Wi^+$; (c) instantaneous volume fraction, $\phi$; (d) transient polymer extensional viscosity, $\eetapp$; symbols are as in Fig.~\ref{f:Rvst-1e-4} (predictions of CDD and FENE-P models are identical in this case).}
\end{figure}

At much larger $\phio \sim \phidagrc$, the friction becomes almost completely Rouse-like and independent of conformation in the CDD model. Predictions at $\phio = 100$ with this model are therefore nearly identical to those of the FENE-P model. The behaviour is qualitatively similar to the FENE-P prediction above for $\phi = 10$, with an inflection at $\Wie = 2/3$ in the SH phase, but with a much more pronounced slow initial linear decay in $R$ with the dominant polymer viscosity close to $\eetapo \approx 3$. There is no hysteresis predicted by the CDD model at this concentration, and $\lame = \lamo = \phio \, \lamZ$ in both models.


\begin{figure}[h!]
\centerline{{\includegraphics{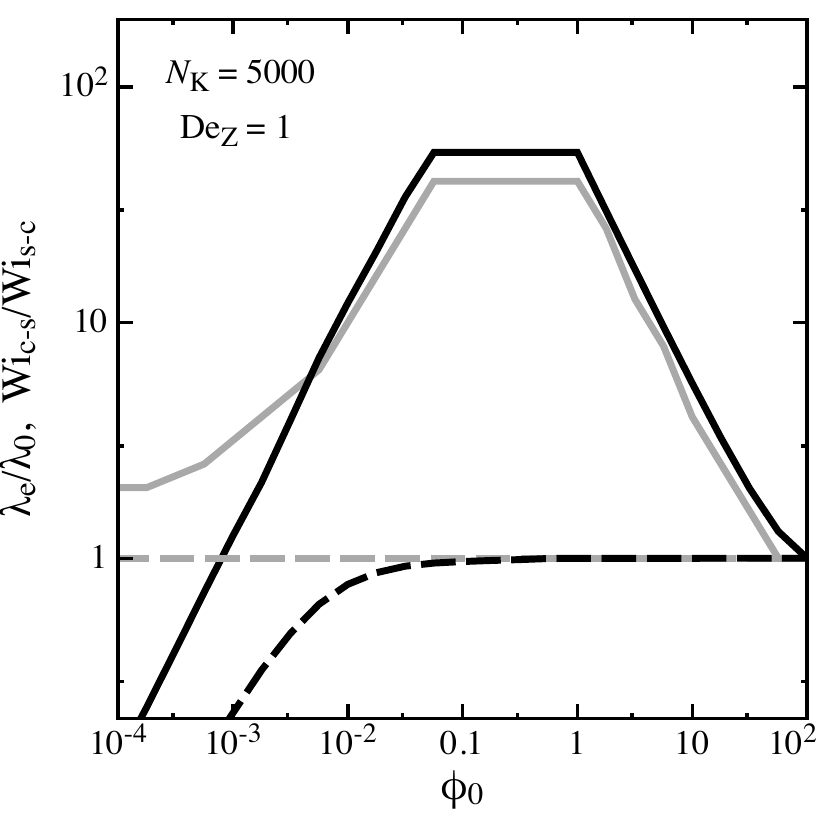}}}
\caption{\label{f:lamebylamo} Comparison of $\lame/ \lamo$ (bold curves) predicted during capillary-thinning with steady-state results for the size of the coil--stretch hysteresis window, $\Wics/\Wisc$ (grey curves): continuous curves - CDD model; dashed curves - FENE-P model.}
\end{figure}

The results above show that beyond a certain concentration, $\Wie$ predicted by the CDD model in capillary-thinning is equal to the critical value $\Wisc$ for the SCT. At all such concentrations $\lame/\lamo$ is therefore proportional to the relative hysteresis window size, since from Eqn.~\eqref{e:lame-wie},
\begin{gather}
\frac{\lame}{\lamo} = \frac{2/3}{\Wisc} = \frac{4}{3}  \,  \frac{\Wics}{\Wisc} \,.
\end{gather}
This is seen clearly in Fig.~\ref{f:lamebylamo} which compares the concentration-dependence of the ratio $\lame/\lamo$ with that of the ratio $\Wics/\Wisc$. Since the FENE-P model does not predict any hysteresis, $\lame/\lamo$ predicted by that model levels off at 1. As mentioned in the Introduction, \citet{prabhakar} also observed that $\Wie = \Wisc$ with a more complex multi-mode model with conformation-dependent intramolecular HI. However, their model did not account for intermolecular interactions; the hysteresis predicted was independent of concentration and much smaller in size, comparable to that observed in single-chain BD results in Fig.~\ref{f:BDScsh}. Their predictions for $\lame/\lamo$ thus also levelled-off with concentration, but to a value larger than unity.

\begin{figure}[h!]
\centerline{{\includegraphics{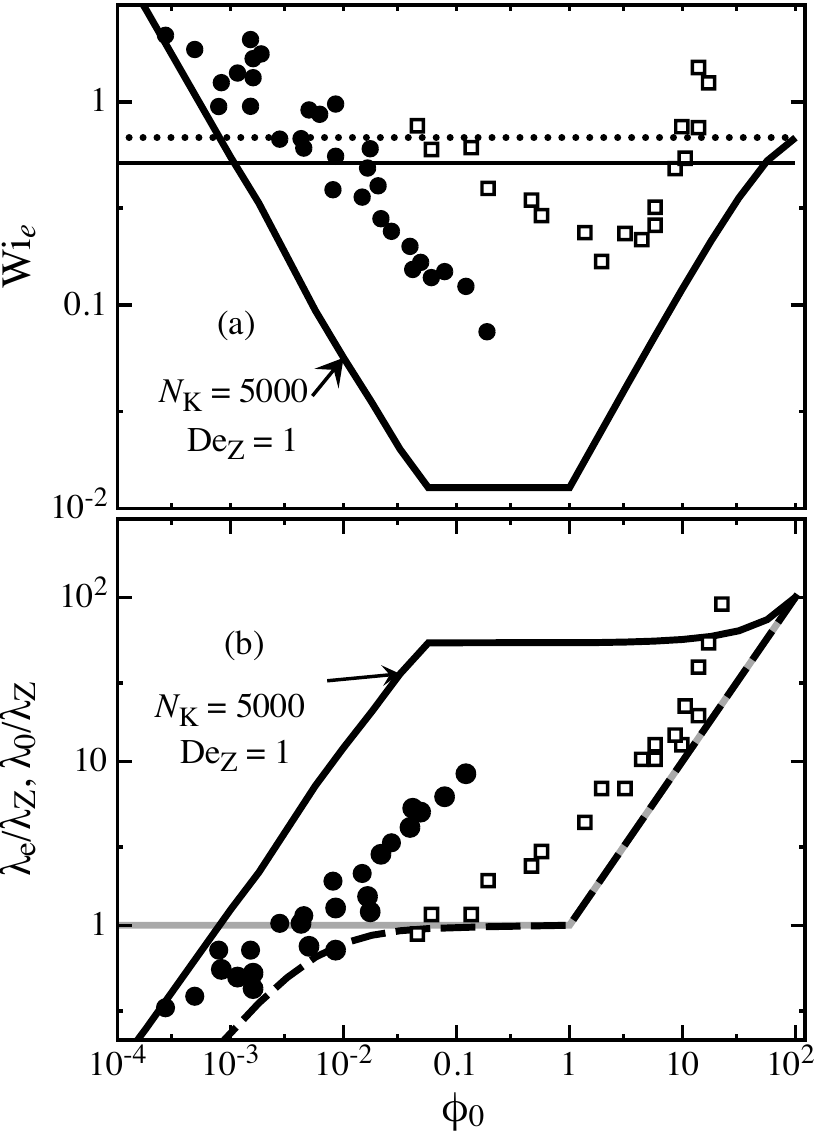}}}
\caption{\label{f:lamebylamz} Qualitative comparison of model predictions with experimental data of \citet{clasenetal} for (a)  $\Wie$ and (b) $\lame/ \lamZ$: filled and open symbols are data for polystyrene solutions in a mixed oligomeric styrene-dioctylphthalate solvent, and in diethylphthalate (DEP), respectively; bold continuous curves are predictions of the CDD model. Model predictions (curves) are all for a fixed value of $\NK = 5000$ and $\Dez = 1$, whereas experimental data are for a range molecular weights (corresponding to $2\times10^3 \lesssim \NK \lesssim 12 \times 10^3$) and Deborah numbers ($0.1 \lesssim \Dez \lesssim 1$).  The continuous and dotted horizontal lines in (a) represent $\Wics =  1/2$ for the coil--stretch transition, and the Entov--Hinch prediction of $\Wie = 2/3$, respectively. In (b), the continuous grey curve is $\lamo/\lamZ = \zetac/\zetaZ$ given by Eqn.~\eqref{e:zetac2}, while the bold dashed line is the prediction of the FENE-P model.}
\end{figure}

Model predictions are compared with the experimental data of \citet{clasenetal} in Fig.~\ref{f:lamebylamz}. The experimental data shown corresponds to polystyrene samples of a range of molecular weights, in two different solvents. The equilibrium relaxation times in the experiments were obtained from SAOS measurements and are described by the empirical Martin equation, $\lamo/\lamZ = \exp ( 0.77 \, K_M\, \phio)$, where $K_M$ is the Martin coefficient. The system-independent universal scaling relationship Eqn.~\eqref{e:zetac2} used for the model predictions gives a concentration dependence that is intermediate between the curves given by the Martin equations for each solvent, with  $K_M = 7.5$ for polystyrene in the oligomeric-styrene Boger fluid, and 0.35 for polystyrene in DEP.  
The effective strain-rate values $\edot_e$ were extracted  by fitting straight lines tangential to  the raw $\log \,R$-vs-$t$ data in the SH phase. The experimental $\Wie = \edot_e \, \lamo$ data shown in Fig.~\ref{f:lamebylamz} (a) have been calculated using those $\edot_e$ values, the $\lamZ$ reported for the different samples by \citeauthor{clasenetal} and  $\lamo$ determined at each experimental $\phio = c/\cstar$ using the empirical Martin equations. In other words, the experimental data are shown ``as-is" from \citet{clasenetal} without using any model equations developed in the present work.

The experimental $\Wie$ data in Fig.~\ref{f:lamebylamz} (a) exhibit two important characteristics. Firstly, as mentioned in the Introduction, it is clear that $\Wie$ during the SH phase fall well below either the Entov-Hinch prediction of $2/3$ or the critical coil--stretch transition $\Wics = 1/2$. Since these data are obtained in the regime when polymer stresses are large and dominant, the data suggest that coil--stretch hysteresis might be important in capillary-thinning dynamics.  Secondly, the trend in the experimental data is in line with the observation of the minimum in $\Wie$ predicted by the CDD model, which results from the minimum in $\Wisc$ with respect to concentration. Figure~\ref{f:lamebylamz} (b) shows the same data in terms of the $\lame/ \lamZ$ ratio, as presented \citeauthor{clasenetal}. Also shown in Fig.~\ref{f:lamebylamz} (b) is the prediction of the conventional FENE-P model. Although the CDD model quite considerably overpredicts the experimental data, the growth in $\lame/\lamZ$ with concentration in the dilute regime is clearly captured by the model. It is possible that the overprediction is due to the neglect of pre-factors in the scaling results used in the model. Taken together with the FENE-P predictions, the model appears to provide bounds on the experimental data. These bounds can further be related to the polymer molecular weight in an unambiguous fashion, and can thus be of significant predictive value in practical applications.  

The CDD model presented here attempts to take the physical insight contained in concepts such as blobs beyond their typical use for explaining scaling exponents and develop a phenomenological but quantitative constitutive model. Nevertheless, it is important to point out the limitations of the modeling presented here. 
\begin{enumerate}
\item Only the slowest mode is considered; extending the model to  develop a multi-mode version requires in principle considering the \textit{dynamic} Zimm-to-Rouse crossover in the relaxation spectrum \citep{ahlrichs2001}.
\item The model neglects the influence of solvent-quality modified excluded-volume (EV) interactions and is restricted to the theta state. The experimental polystyrene systems considered by \citet{clasenetal} show a significant influence of  such interactions. Incorporating those effects require considering ``theta blobs" and EV screening when partially stretched chains overlap. In addition, EV interactions also influence the stiffness of the entropic resistance \citep{pincus1976}, and therefore affect the coil--stretch transition and hysteresis \citep{radhaunderhill, shika}.
\item The influence of entanglements are completely ignored, which are known to become important at a concentration $\phio \sim 10$; the results presented above consider $\phio$ values well above that limit. An interesting question in this context is whether chains can become dynamically entangled due to self-concentration. Incorporating such effects into a CDD model at the dumbbell level might be possible through the ``encapsulated-dumbbell" model \citep{birddeaguiar, fang}.
\item It must be pointed out here that although the values $\NK$ and $\Dez$ used in the modeling are comparable with the experimental values, the Ohnesorge number $\Oh = \etas/\sqrt{\rho\,\gamma \, R_0}$ (where $\rho$ is the solution density) varies considerably across the two solvents. This number quantifies the relative importance of viscous effects over inertio-capillary effects. The stress-balance considered in the present study neglects fluid inertia, and as such the results obtained here are valid only in the limit of $\Oh \rightarrow \infty$. Among the two experimental systems of \citeauthor{clasenetal}, the polystyrene Boger fluid has $\Oh \sim 10^2$ for which the inertialess stress-balance is appropriate. The polystyrene-in-DEP system on the other hand has $\Oh \sim 10^{-3}$. Modeling systems of moderate or low $\Oh$ more accurately requires generalizing the stress-balance with an inertial term, as suggested by \citet{viyaada}.
\end{enumerate}

The results above suggest that capillary-breakup and indeed filament-stretching extensional rheometers devices could play a vital role in discriminating  between theories and models for self-concentration and thus contribute to an improved understanding of the dynamics of polymer solutions, particularly coil--stretch hysteresis. Although proposed in the 1970s, to this author's best knowledge, there has been thus far only \textit{one} other rheological observation of hysteresis in filament-stretching rheometry \cite{sridharPRL}, which together with the single-molecule studies of \citet{schroeder}, constitute the sum total  of experimental evidence for this phenomenon. While the practical importance of the coil--stretch \textit{transition} is widely appreciated, the data in Fig.~\ref{f:lamebylamz} show that hysteresis may also play a vital role in a number of applications where capillary-thinning is relevant. The present work thus points to the need for more systematic experiments to explore the role played by parameters such as $\NK$, $\Dez$, $\Oh$ and solvent quality on capillary thinning. On the modeling front, the results here underline the necessity for detailed multi-chain BD simulations and molecular theory to  develop models incorporating conformation-dependent intra- and inter-molecular HI. 

It is possible that the influence of self-concentration will not be as dramatic as that observed above in all kinds of flows. The effect stems from the combination of high stretching and low strain-rates that is peculiar to the hysteretic domain in extensional flow. A similar strong effect is unlikely at steady-states in shear flow, where high stretching also requires high strain-rates, which dampen transverse fluctuations. On the other hand, self-concentration could be significant during relaxation of initially stretched chains to equilibrium after cessation of either shear or extensional flow, since transverse fluctuations and polymer stretch can be simultaneously large under those conditions.

\section{\label{s:concl} Conclusions}
The principal conclusion of this work is that the anomalous concentration dependence observed in capillary-thinning experiments and multi-chain BD simulations on dilute and semidilute polymer solutions might be explained by an increased hydrodynamic interaction between chain molecules that pervaded larger volumes in flow situations that cause significant chain stretching but do not dampen transverse conformational fluctuations. Such self-concentration could lead to a strong enhancement of the phenomenon of coil--stretch hysteresis in nominally dilute polymer solutions. The enhancement is maximal over a range of concentrations well below the $c/\cstar = 1$, and progressively diminishes in the semidilute regime.

\acknowledgments{This work was supported by a CPU-time grant on the National Computational Infrastructure at the Australian National University, Canberra, and a monetary grant from the Australian Research Council (Discovery Project No. DP120101322). The author is indebted to Ravi Jagadeeshan, Burkhard Duenweg,  Gareth McKinley, Tam Sridhar and David Boger for their insight, advice and encouragement.}

\appendix

\section{\label{a:stretch} Analysis of the stable stretched state in steady uniaxial extensional flow}
The qualitative features of coil--stretch hysteresis in the dilute limit have been in the past explained by considering the the steady-state probability distribution of  end-to-end stretch in uniaxial extensional flow of a dumbbell with conformation-dependent friction \citep{degennes, schroeder_science, schroeder}. The CDD model presented in the current work also captures many of the important features. 

At steady state for a give extension rate $\edot$, 
\begin{eqnarray}
0 = 2 \,\edot\, \ell^2 - \frac{1}{\lamo \, (\zeta/\zeta_0)}\, \left( \,f\, \ell^2- d_0^2 \right) \,,\\[3mm] \label{e:mxxodess}
0 = - \edot\, d^2 - \,\frac{1}{\lamo \, (\zeta/\zeta_0)}\, \left( \,f\, d^2 - d_0^2 \right)\,.
\label{e:myyodess}
\end{eqnarray}
From the exact results, one expects that in the stable stretched state (denoted by the subscript `$s$'), $\fs\,\ells^2 \gg  d_0^2 \gtrsim d^2$. Substituting this approximation in Eqn.~\eqref{e:mxxodess} above,
\begin{gather}
\fs \,= \,2\, \Wi \, \frac{\zetas}{\zeta_0} \,.
\label{e:fs1}
\end{gather}
Assuming that $\zetas \approx \zetar$ in the stretched state, from the equations for $\zetar/\zetaZ$ in Table~\ref{t:drageqns} and Eqn.~\eqref{e:zetac2}, $\zetas/\zetac$ is of the form $\zetas/\zetac = C \, \ells/d_0$. In general, $C$ is itself a function of the conformation, but in situations when $C$ is nearly constant, an approximate solution can be obtained. This occurs in the dilute limit, and when HI screening is close to complete in the stretched state and $\zetar \approx \zetaR$. Formally, a detailed analysis of the stretched state requires a perturbation expansion of all terms depending on $\ells$ as $\ells = L( 1 - \epsilon + \ldots)$. For the sake of clarity, $C$ is assumed to be constant below, which yields qualitatively similar results although with different pre-factors. 
\begin{gather}
\fs \,= 2\,C \,\Wi \,\frac{\ells}{d_0}\,.
\label{e:fs2}
\end{gather}
From the definition of $f$ in Eqn~\eqref{e:fdef}, in the stretched state
\begin{gather}
\fs \approx \frac{L^2}{L^2 - \ells^2} \approx \frac{L}{2\, (L - \ells)} \,.
\label{e:fs3}
\end{gather}
Equating the two expressions above for $\fs$ gives a quadratic equation for $\ells/L$:
\begin{gather}
\frac{\ells}{L} \, \left( 1- \frac{\ells}{L} \right) - \frac{d_0}{4 \, \Wi \, C\, L} = 0 \,,
\label{e:quadeqn}
\end{gather}
which has real roots when $\Wi$ is greater than the critical value
\begin{gather}
\Wisc = \frac{d_0}{C \, L} = \frac{1}{C \, \sqrt{3 \, \NK}} \,,
\label{e:C-Wisc}
\end{gather}
since $L/d_0 = \sqrt{3 \,\NK}$.

Thus, the constant $C$ can be eliminated in favour of $\Wisc$. For $\Wi > \Wisc$, $\ells$ for the stable stretched state corresponds is given by the root of the quadratic equation above:
\begin{gather}
\ells = \frac{L}{2} \, \left[1 + \left(1 - \frac{\Wisc}{\Wi}\right)^{1/2} \right]\,,
\label{e:ls}
\end{gather}
and $\ells$ approaches $L$  with increasing $\Wi$. Substituting for $C$ and $\ells$ in Eqn.~\eqref{e:fs2}, 
\begin{gather}
\fs = \frac{\Wi}{\Wisc}\,\left[1 + \left(1 - \frac{\Wisc}{\Wi}\right)^{1/2} \right] \,.
\label{e:fs4}
\end{gather}
Considering next Eqn.~\eqref{e:myyodess} for the transverse size $d_\mathrm{s}$ in the stretched state,
\begin{gather}
\left(\frac{d_\mathrm{s}}{d_0}\right)^2 = \frac{1}{\fs + \Wi \,(\zeta_\mathrm{s}/\zeta_0)} =  \frac{1}{3 \, \Wi \, (\zeta_\mathrm{s}/\zeta_0)} = \frac{2}{3\,\fs} = \frac{2\, \Wisc}{3\, \Wi\,\left[1 + \left(1 - (\Wisc/\Wi)\right)^{1/2} \right]} \,.
\label{e:dsbyd0}
\end{gather}
The pervaded volume  $V = \ell\, d^2$;  in the stretched state therefore, from Eqns.~\eqref{e:ls} and \eqref{e:dsbyd0} above, 
\begin{gather}
\frac{\phi_\mathrm{s}}{\phio} = \frac{V_\mathrm{s}}{V_0} = \left(\frac{\ells}{d_0} \right)\, \left(\frac{d_\mathrm{s}}{d_0} \right)^2 = \frac{1}{3} \,\frac{L/d_0}{\Wi/\Wisc}  = \frac{\sqrt{\NK/3}}{\Wi/\Wisc}\,  \,.
\label{e:VsbyV0}
\end{gather}
Therefore, transverse chain dimensions and volume are largest at the SCT and decrease with increasing strain-rate  effectively as $\Wi^{-1}$.

Chain dimensions at the stable stretched state just at the SCT are obtained by setting $\Wi = \Wisc$ in the relations above. This state is referred to as the $\Upsigma$-state, and values therein are denoted with the subscript $\Upsigma$. Thus, from Eqn.~\eqref{e:ls},
\begin{gather}
\frac{\ell_\Upsigma}{d_0} = \frac{L}{2\, d_0} = \frac{\sqrt{3\,\NK}}{2} \,,\label{e:ellsc} \\
\left(\frac{d_\Upsigma}{d_0}\right)^2= \frac{2}{3}\,,\label{e:dsc} \\
\frac{V_\Upsigma}{V_0} = \frac{\phi_\Sigma}{\phio} = \sqrt{\frac{\NK}{3}} \label{e:Vsc} \,.
\end{gather} 
Further, from Eqn.~\eqref{e:fs4}, $f_\Upsigma = 1$ in the $\Upsigma$-state; therefore, from Eqn.~\eqref{e:fs1},
\begin{gather}
2\, \Wisc \, \frac{\zeta_\Upsigma}{\zetac} = 1 \,,\notag \\
\intertext{or equivalently,}
\edot_\Upsigma \,\left( \lamo \, \frac{\zeta_\Upsigma}{\zetac} \right) = \edot_\Upsigma \,\lambda_\Upsigma = \frac{1}{2} \,.
\end{gather}
The equation above shows that the SCT occurs when a Weissenberg number defined with a polymer relaxation-time characteristic of the  stretched state---$\lambda_\Upsigma = \lamo (\zeta_\Upsigma/\zetac)$---attains a value of $1/2$. Noting that $\Wics = 1/2$, the width of the hysteresis window is obtained as
\begin{gather}
\frac{\Wics}{\Wisc}\, =  \, \frac{\zeta_\Upsigma}{\zetac} \,.
\label{e:wiratio-zetaratio-a}
\end{gather}

\end{document}